\documentclass[twocolumn, superscriptaddress, amsmath,amssymb, prd, aps, longbibliography]{revtex4-1}

\usepackage{romanbar}
\usepackage{xcolor}
\usepackage{graphicx}% Include figure files
\usepackage{dcolumn}% Align table columns on decimal point
\usepackage{bm}% bold math
\usepackage{SIunits}
\usepackage{array}
\usepackage{subfigure}
\usepackage{multirow}%

\usepackage{hyperref} 

\usepackage{url}

\begin{document}

%\title{Performance of the COSINE-100U experiment}
\title{Upgrading the COSINE-100 Experiment for Enhanced Sensitivity to  Low-Mass Dark Matter Detection}
\author{D.~H.~Lee}
\affiliation{Department of Physics, Kyungpook National University, Daegu 41566, Republic of Korea}
\author{J.~Y.~Cho}
\affiliation{Department of Physics, Kyungpook National University, Daegu 41566, Republic of Korea}
\author{C.~Ha}
\affiliation{Department of Physics, Chung-Ang University, Seoul 06973, Republic of Korea}
\author{E.~J.~Jeon}
\affiliation{Center for Underground Physics, Institute for Basic Science (IBS), Daejeon 34126, Republic of Korea}
\affiliation{IBS School, University of Science and Technology (UST), Daejeon 34113, Republic of Korea}
\author{H.~J.~Kim}
\affiliation{Department of Physics, Kyungpook National University, Daegu 41566, Republic of Korea}
\author{J.~Kim}
\affiliation{Department of Physics, Chung-Ang University, Seoul 06973, Republic of Korea}
\author{K.~W.~Kim}
\affiliation{Center for Underground Physics, Institute for Basic Science (IBS), Daejeon 34126, Republic of Korea}
\author{S.~H.~Kim}
\affiliation{Center for Underground Physics, Institute for Basic Science (IBS), Daejeon 34126, Republic of Korea}
\author{S.~K.~Kim}
\affiliation{Department of Physics and Astronomy, Seoul National University, Seoul 08826, Republic of Korea}
\author{W.~K.~Kim}
\affiliation{IBS School, University of Science and Technology (UST), Daejeon 34113, Republic of Korea}
\affiliation{Center for Underground Physics, Institute for Basic Science (IBS), Daejeon 34126, Republic of Korea}
\author{Y.~D.~Kim}
\affiliation{Center for Underground Physics, Institute for Basic Science (IBS), Daejeon 34126, Republic of Korea}
\affiliation{IBS School, University of Science and Technology (UST), Daejeon 34113, Republic of Korea}
\author{Y.~J.~Ko}
\affiliation{Department of Physics, Jeju National University, Jeju 64243, Republic of Korea}
\author{H.~Lee}
\affiliation{IBS School, University of Science and Technology (UST), Daejeon 34113, Republic of Korea}
\affiliation{Center for Underground Physics, Institute for Basic Science (IBS), Daejeon 34126, Republic of Korea}
\author{H.~S.~Lee}
\email{hyunsulee@ibs.re.kr}
\affiliation{Center for Underground Physics, Institute for Basic Science (IBS), Daejeon 34126, Republic of Korea}
\affiliation{IBS School, University of Science and Technology (UST), Daejeon 34113, Republic of Korea}
\author{I.~S.~Lee}
\email{islee@ibs.re.kr}
\affiliation{Center for Underground Physics, Institute for Basic Science (IBS), Daejeon 34126, Republic of Korea}
\author{J.~Lee}
\affiliation{Center for Underground Physics, Institute for Basic Science (IBS), Daejeon 34126, Republic of Korea}
\author{S.~H.~Lee}
\affiliation{IBS School, University of Science and Technology (UST), Daejeon 34113, Republic of Korea}
\affiliation{Center for Underground Physics, Institute for Basic Science (IBS), Daejeon 34126, Republic of Korea}
\author{S.~M.~Lee}
\affiliation{Department of Physics and Astronomy, Seoul National University, Seoul 08826, Republic of Korea} 
\author{R.~H.~Maruyama}
\affiliation{Department of Physics and Wright Laboratory, Yale University, New Haven, CT 06520, USA}
\author{J.~C.~Park}
\affiliation{Department of Physics and IQS, Chungnam National University, Daejeon 34134, Republic of Korea}
\author{K.~S.~Park}
\affiliation{Center for Underground Physics, Institute for Basic Science (IBS), Daejeon 34126, Republic of Korea}
\author{K.~Park}
\affiliation{Center for Underground Physics, Institute for Basic Science (IBS), Daejeon 34126, Republic of Korea}
\author{S.~D.~Park}
\affiliation{Department of Physics, Kyungpook National University, Daegu 41566, Republic of Korea}
\author{K.~M.~Seo}
\affiliation{Center for Underground Physics, Institute for Basic Science (IBS), Daejeon 34126, Republic of Korea}
\author{M.~K.~Son}
\affiliation{Department of Physics and IQS, Chungnam National University, Daejeon 34134, Republic of Korea}
\author{G.~H.~Yu}
\affiliation{Department of Physics, Sungkyunkwan University, Suwon 16419, Republic of Korea}
\affiliation{Center for Underground Physics, Institute for Basic Science (IBS), Daejeon 34126, Republic of Korea}

\begin{abstract}
The DAMA/LIBRA experiment has reported an annual modulation signal in NaI(Tl) detectors, which has been interpreted as a possible indication of dark matter interactions. However, this claim remains controversial, as several experiments have tested the modulation signal using NaI(Tl) detectors. Among them, the COSINE-100 experiment, specifically designed to test DAMA/LIBRA's claim, observed no significant signal, revealing a more than 3$\sigma$ discrepancy with DAMA/LIBRA's results. Here we present COSINE-100U, an upgraded version of the experiment, which aims to expand the search for dark matter interactions by improving light collection efficiency and reducing background noise. The detector, consisting of eight NaI(Tl) crystals with a total mass of 99.1 kg, has been relocated to Yemilab, a new underground facility in Korea, and features direct PMT-coupling technology to enhance sensitivity. These upgrades significantly improve the experiment's ability to probe low-mass dark matter candidates, contributing to the ongoing global effort to clarify the nature of dark matter.
\end{abstract} 

%\keywords{Dark Matter, COSINE-100U, NaI(Tl)}

\maketitle

\section{Introduction}\label{sec1}
Numerous astronomical observations suggest that the majority of matter in the universe consists of invisible dark matter, though its nature and interactions remain elusive~\cite{Clowe:2006eq,Planck:2018vyg, Bertone:2016nfn}. Despite extensive efforts to directly detect dark matter, no definitive signals have been observed~\cite{MarrodanUndagoitia:2015veg, Schumann:2019eaa}. The only exception is the DAMA/LIBRA experiment, which has reported annual modulation signals in a 250\,kg array of NaI(Tl) detectors~\cite{Bernabei:1998fta,Bernabei:2013xsa,Bernabei:2018yyw,Bernabei:2021kdo}, indicating potential dark matter-nuclei interactions~\cite{Savage:2008er, COSINE-100:2019brm}. 
However, these results remain controversial, as no other experiments have observed similar signals~\cite{Schumann:2019eaa, ParticleDataGroup:2022pth}, necessitating independent verification using the same NaI(Tl) crystal target materials.

Several experimental efforts are currently underway to replicate these findings using NaI(Tl) as the target medium~\cite{Adhikari:2017esn,Amare:2018sxx,Fushimi:2021mez,Antonello:2020xhj,SABRE:2022twu,COSINUS:2023kqd}. 
Among these, the COSINE-100 experiment, which began operation in October 2016 at the Yangyang Underground Laboratory (Y2L) with 106\,kg of NaI(Tl) crystals, was the first to follow DAMA/LIBRA~\cite{Adhikari:2017esn}. 
COSINE-100 aimed to test the DAMA/LIBRA findings, and data from the experiment generally align with null results for WIMP-nuclei interactions~\cite{Adhikari:2018ljm,COSINE-100:2021xqn}, challenging the interpretation of the DAMA/LIBRA signal under standard weakly interacting massive particle (WIMP) dark matter models. Additionally, model-independent annual modulation searches from COSINE-100 show a more than 3$\sigma$ discrepancy with the DAMA/LIBRA results~\cite{COSINE-100:2024jkg}. Further studies are needed to fully understand DAMA/LIBRA's results; however, improving detector performance, such as lowering the analysis threshold and reducing internal background contamination, is essential for enhancing dark matter detection sensitivities for the next phase of experiments.

The COSINE-100 experiment concluded in March 2023 after 6.4\,years of stable operation. It was relocated to the newly constructed Yemilab~\cite{Park:2024sio,Yemilab2024} in Korea for the next phase, COSINE-100U. During the transition, the crystal encapsulation was upgraded to improve light collection by directly attaching the crystals to photomultiplier tubes~(PMTs)~\cite{Choi:2020qcj}. A similar technique, applied in the NEON experiment for measuring coherent elastic neutrino-nucleus scattering at nuclear reactors~\cite{NEON:2022hbk}, achieved approximately a 50\% improvement in light yield and demonstrated long-term stability over two years~\cite{Choi:2024trx}. 
These improvements in light yield are expected to lower the energy threshold, thereby improving the sensitivity of COSINE-100U to dark matter, particularly in the low-mass region. For the COSINE-100U experiment at Yemilab, all eight NaI(Tl) crystals of COSINE-100 were upgraded. In this article, we detail the detector upgrade and the resulting improved sensitivities for low-mass dark matter detection.

\section{NaI(Tl) Crystal Encapsulation}\label{sec2}
The COSINE-100 experiment~\cite{Adhikari:2017esn} was designed to investigate the DAMA/LIBRA annual modulation claim~\cite{Bernabei:2021kdo} using low-background NaI(Tl) detectors. The detector consists of eight NaI(Tl) crystals, each hermetically encapsulated in oxygen-free copper (OFC) enclosures and immersed in a liquid scintillator veto system~\cite{Adhikari:2020asl} to suppress external backgrounds. While COSINE-100 has successfully operated for over six years, improvements in light collection efficiency are necessary for the next phase, COSINE-100U, to enhance sensitivity to low-energy dark matter interactions.

One of the key upgrades in COSINE-100U involves a new crystal encapsulation technique to address the limitations of the original COSINE-100 design. The quartz windows used in the original setup introduced additional photon loss due to multiple optical interfaces between the optical pad, quartz window, and optical grease. The light collection efficiency was further limited by the mismatch between the crystal diameter and the PMTs, requiring beveled edges to guide photons toward the 3-inch PMTs. To improve light yield and signal quality, the COSINE-100U design eliminates the quartz windows and adopts a direct PMT-coupling approach, where the PMTs are attached directly to the crystal using a 2 mm thick optical pad. This method, initially demonstrated in the NEON experiment~\cite{Choi:2020qcj,Choi:2024trx}, has been shown to increase light yield by approximately 50\%, reaching approximately 22 photoelectrons (NPE)/keV, while maintaining long-term stability over two years.

NaI(Tl) crystals are typically packaged by commercial manufacurers using aluminum or copper enclosures with quartz windows, which protect the material while maintaining optical transparency. To enhance light collection efficiency, the crystal surfaces are often wrapped in reflective materials, such as Teflon sheets or aluminum oxide powder, which help maximize photon reflection. The COSINE-100 experiment utilized eight NaI(Tl) crystals grown by Alpha Spectra Inc., produced in collaboration with the KIMS, DM-Ice, and ANAIS collaborations to ensure low-background purity~\cite{Kim:2014toa,Amare:2014jta,Adhikari:2015rba}.

\begin{figure}[!htb]
    \includegraphics[width=1.0\columnwidth]{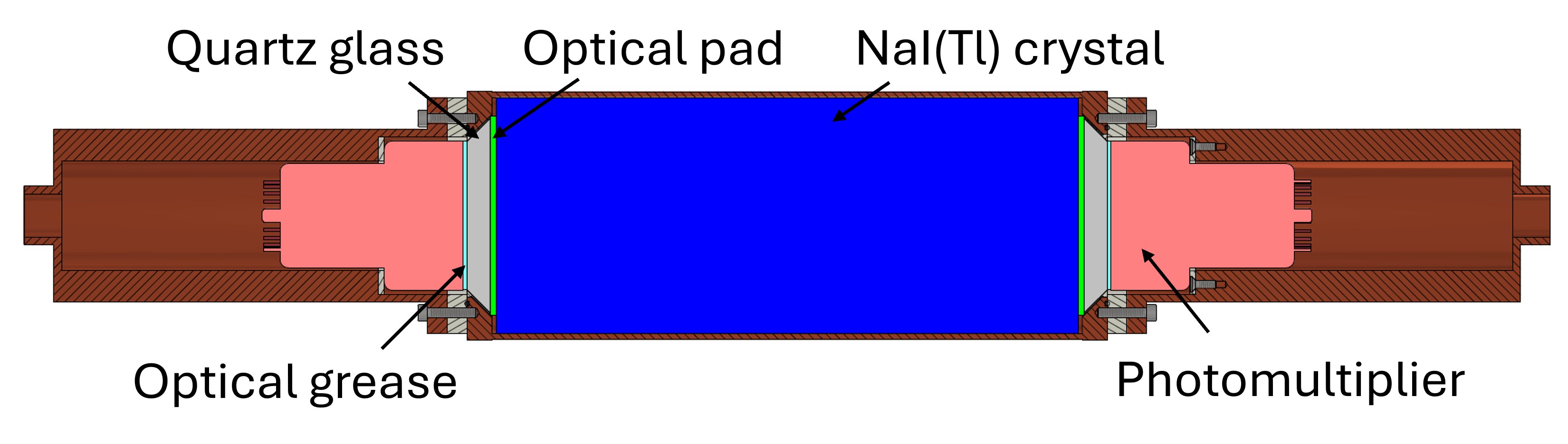}
  \caption{
{\bf The encapsulation design of the COSINE-100 C6 crystal.} 
A 4.8-inch diameter crystal is encased in OFC tubes, with an optical pad and quartz windows. The crystal shown in the figure is wrapped with a Teflon sheet.
The 12\,mm thick quartz window guides light from the 4.8-inch diameter crystal to the 3-inch PMTs, aided by a 45$^{\circ}$ angle surface. There are three optical interface layers between the crystal and the PMTs: a 1.5\,mm optical pad, a 12\,mm quartz windows, and a small amount of optical grease.  The PMTs are encapsulated in OFC cases to protect them from the surrounding liquid scintillator. 
    }
\label{fig:cosine100-encap}
\end{figure}

Each of these cylindrical crystals, with slightly varying dimensions as listed in Table~\ref{table:Crystals}, was hermetically encased in OFC tubes (1.5\,mm thick) with quartz windows at both ends, as exemplified by crystal-6 (C6) in Fig.~\ref{fig:cosine100-encap}. The lateral surfaces of each crystal were wrapped in Teflon sheets before being inserted into the OFC tubes. A 12\,mm thick quartz window was coupled to the crystal using a 1.5\,mm thick optical pad, and PMTs were attached to the quartz window using a small amount of high-viscosity optical grease.

From the perspective of light collection efficiency, the original COSINE-100 encapsulation design had several drawbacks. The three-layer optical interface, comprising the optical pad, quartz window, and optical grease, introduced additional reflections, reducing light collection efficiency. Although the 12\,mm thick quartz window, with a 45$^{\circ}$ angle, guided photons from the 4.8-inch diameter crystal, it was insufficient for efficiently directing photons to the 3-inch PMTs. As shown in Fig.~\ref{fig:cosine100-encap}, uncovered areas resulted in photon reflection, further reducing light collection efficiency.

To address these issues, an enhanced crystal encapsulation technique was developed, initially in the NEON experiment, which involved directly attaching the PMT to the crystal using only a 2\,mm thick optical pad~\cite{Choi:2020qcj}. In this method, the crystal's diameter was matched to the 3-inch diameter of the PMTs, eliminating photon absorption in the quartz and minimized photon loss due to reflections from multiple interfaces. This modification resulted in a light yield of up to 22 NPE/keV, approximately 50\% higher than the light yield observed in COSINE-100 crystals, which are measured around 15\,NPE/keV~\cite{Adhikari:2017esn}. 

This technique was initially applied in the NEON experiment during an engineering run at a nuclear reactor to observe coherent elastic neutrino-nucleus scattering~\cite{NEON:2022hbk}. However, some design weaknesses were identified, as liquid scintillator leakage into the detector caused a gradual decrease in crystal light yield and increase of PMT-induced noise.

%Since then, we have improved the crystal encapsulation by separating it into two components: an inner structure to maintain a stable coupling between the PMTs and crystal, and an outer OFC case to prevent the infiltration of external air and liquid scintillator. These improvements were successfully applied to the NEON experiment, which has been collecting stable physics data for over two years with above 22\,NPE/keV high light yield~\cite{Choi:2024trx}. 

The updated design separates the encapsulation into two components: an inner structure to maintain a stable coupling between the PMTs and crystal, and an outer OFC case to prevent the infiltration of external air and liquid scintillator. These improvements were successfully applied to the NEON experiment, which has been collecting stable physics data for over two years with a light yield exceeding 22\,NPE/keV~\cite{Choi:2024trx}. 

\begin{figure}[!htb]
  \begin{center}
    \includegraphics[width=1.0\columnwidth]{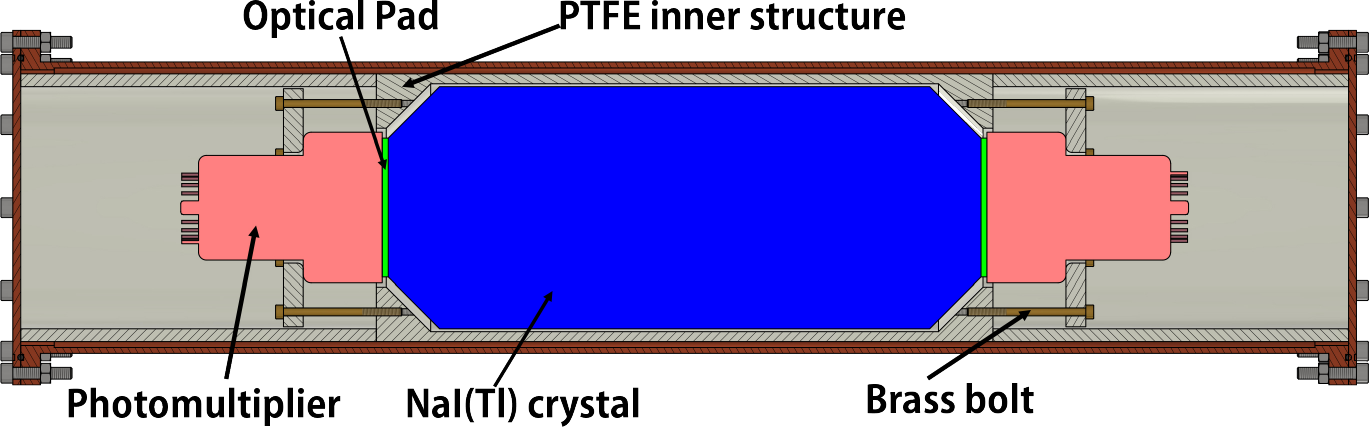}
  \end{center}
  \caption{
  {\bf The encapsulation design of the COSINE-100U C6 crystal.} 
The edges of a 4.8-inch diameter crystal are beveled at a 45$^{\circ}$ angle to guide light to 3-inch PMTs and wrapped with a Teflon sheet. PMTs are directly attached to the crystal using a 2 mm thick optical pad and supported by a rigid 5 mm thick PTFE inner structure. The crystal-PMT assembly, along with the inner structure, is encased in OFC tube to protect it from the surrounding liquid scintillator. 
}
\label{fig:cosine100U-encap}
\end{figure}

Based on this experience, some of the current authors, who were also involved in the NEON experiment, applied the lessons learned to improve the crystal encapsulation for COSINE-100U,
 as shown in Fig.~\ref{fig:cosine100U-encap}. While the NEON crystals have a 3-inch diameter matching the size of 3-inch PMTs, the COSINE-100 crystals have larger diameters, as listed in Table~\ref{table:Crystals}. To address this, the edges of the COSINE-100 crystals are beveled at a 45$^{\circ}$ angle, reducing the edge diameter to 3-inch and effectively guiding light from the larger-diameter crystals to the 3-inch PMTs.

To ensure stable mounting of the PMTs to the crystals, a 5\,mm thick PTFE inner structure is employed, which fully encases the crystal. A 2\,mm thick optical pad is positioned between the PMTs and the crystal surface to optimize light transmission. Stable light coupling is achieved through a PTFE ring, secured with brass bolts to the PTFE inner structure, which applies consistent and uniform pressure to the optical pad. 
The assembled crystals are hermetically encased in 2\,mm thick OFC tubes, with both ends sealed by 20\,mm thick OFC lid flanges.  Further details regarding the COSINE-100U crystal assembly process can be found in Appendix~\ref{sec8}.

\begin{table}
\centering
\caption{{\bf Summary of the COSINE-100 and COSINE-100U crystals geometry.} The COSINE-100 experiment originally consisted of eight NaI(Tl) crystals with a total mass of 106.3\,kg. However, three crystals (marked with an asterisk) were excluded from the dark matter search, resulting in an effective mass of 61.4\,kg. In the COSINE-100U, all eight crystals were included in the analysis, but the total mass was reduced to 99.1\,kg due to the machining process.}
\begin{tabular}{ c c  c  c }
\hline
\multirow{2}{*}{Crystal} & Size (inches) & \multicolumn{2}{c}{Mass [kg]} \\
    & diameter$\times$length &  COSINE-100  &  COSINE-100U   \\\hline\hline
C1 & 5.0$\times$7.0  & 8.3$^*$  & 7.1   \\ 
C2 & 4.2$\times$11.0 & 9.2  & 8.7   \\ 
C3 & 4.2$\times$11.0 & 9.2  & 8.7   \\ 
C4 & 5.0$\times$15.3 & 18.0 & 16.9  \\ 
C5 & 5.0$\times$15.5 & 18.3$^*$ & 17.2  \\ 
C6 & 4.8$\times$11.8 & 12.5 & 11.7  \\ 
C7 & 4.8$\times$11.8 & 12.5 & 11.6  \\ 
C8 & 5.0$\times$15.5 & 18.3$^*$ & 17.2  \\ \hline
Total &  & 106.3 (61.4) & 99.1   \\ \hline
\end{tabular}
\label{table:Crystals}
\end{table}

\section{Sea Level Measurements}\label{sec3}
\subsection{Measurement Setup}
\begin{figure}[tb!]
    \centering
    \includegraphics[width=0.59\columnwidth]{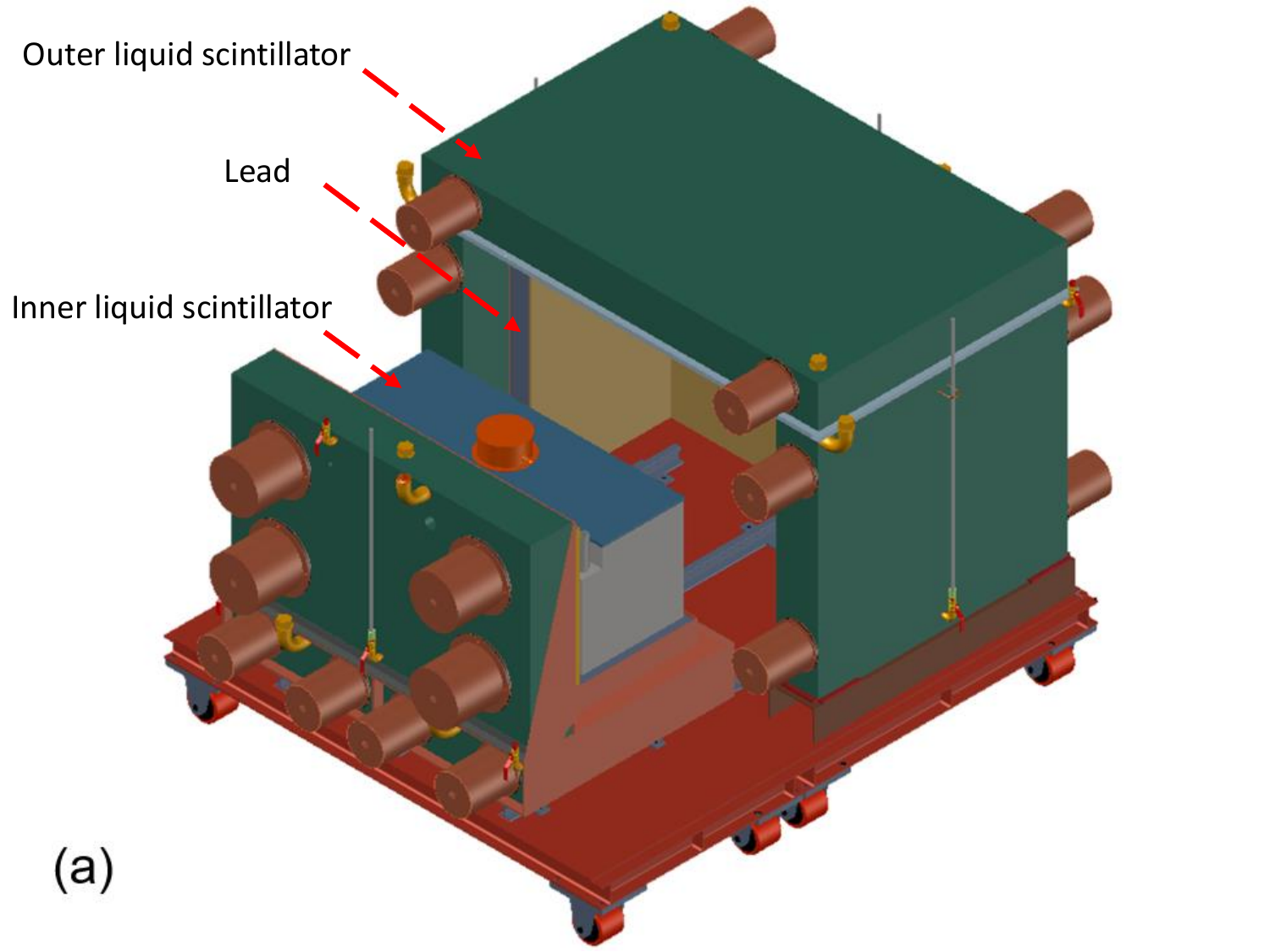}    \includegraphics[width=0.39\columnwidth]{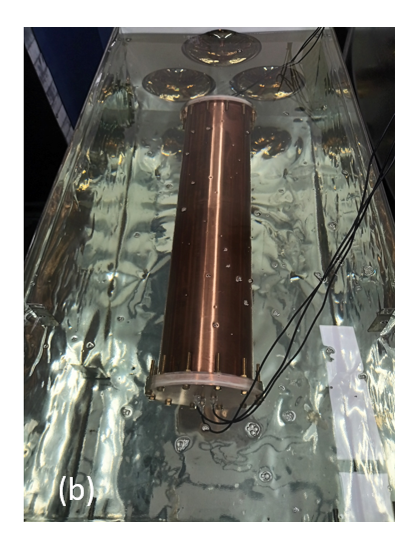}
    \caption{
    {\bf Test measurements setup.}
    (a)  
    Schematic view of the shield structure used for the sea-level measurement at the Institute for Basic Science. The encapsulated crystal detector is installed within the inner liquid scintillator layer, approximately 20 cm thick. This layer is surrounded by a 10 cm thick lead shield and an outer liquid scintillator container. For this measurement, the outer liquid scintillator container was not filled.
    (b) The C1 crystal is installed in this setup for light yield and stability measurements. }
    \label{fig:testbench}
\end{figure}

Upon assembling the COSINE-100 crystals, the Yemilab facility was not yet ready for operation of the COSINE-100U experiment. For the initial crystal characterization, we employed a simple shielding setup at sea level in the experimental hall of the Institute for Basic Science (IBS) in Korea. This setup consisted of two layers of shielding: 10\,cm thick lead and 20\,cm thick liquid scintillator, which also functioned as an active veto detector. The schematic view of the shielding structure is shown in Fig.~\ref{fig:testbench}(a). The liquid scintillator was housed within a 124.5\,cm $\times$ 49.5\,cm $\times$ 49.5\,cm cubic stainless steel box. Three 8-inch PMTs were used to read the signals from the liquid scintillator as shown in Fig.~\ref{fig:testbench}(b). This setup was initially developed as a prototype detector for the NEOS experiment~\cite{NEOS:2016wee} and was reused for this test.  Inside the container, an acrylic table was used to install one NaI(Tl) crystal for the initial evaluation of its performance, as shown in Fig.~\ref{fig:testbench}(b). 

Events were collected by two PMTs attached to each crystal, with the photoelectron signals amplified using a preamplifier and digitized by a flash analog-to-digital converter (FADC) at a sampling rate of 500\,MHz. The FADC, connected to the trigger control board (TCB), recorded events that satisfied specific trigger conditions, capturing waveforms over an 8\,$\mu$s window starting 2.4\,$\mu$s before the trigger time~\cite{COSINE-100:2018rxe}. Additionally, signals from three 8-inch PMTs for the liquid scintillator were recorded using another 500\,MHz FADC with the same 8\,$\mu$s recording length when NaI(Tl) crystal-triggered events occurred.

\subsection{Light Yield Measurement}
\begin{figure}[tb!]
    \centering
    \includegraphics[width=1.0\columnwidth]{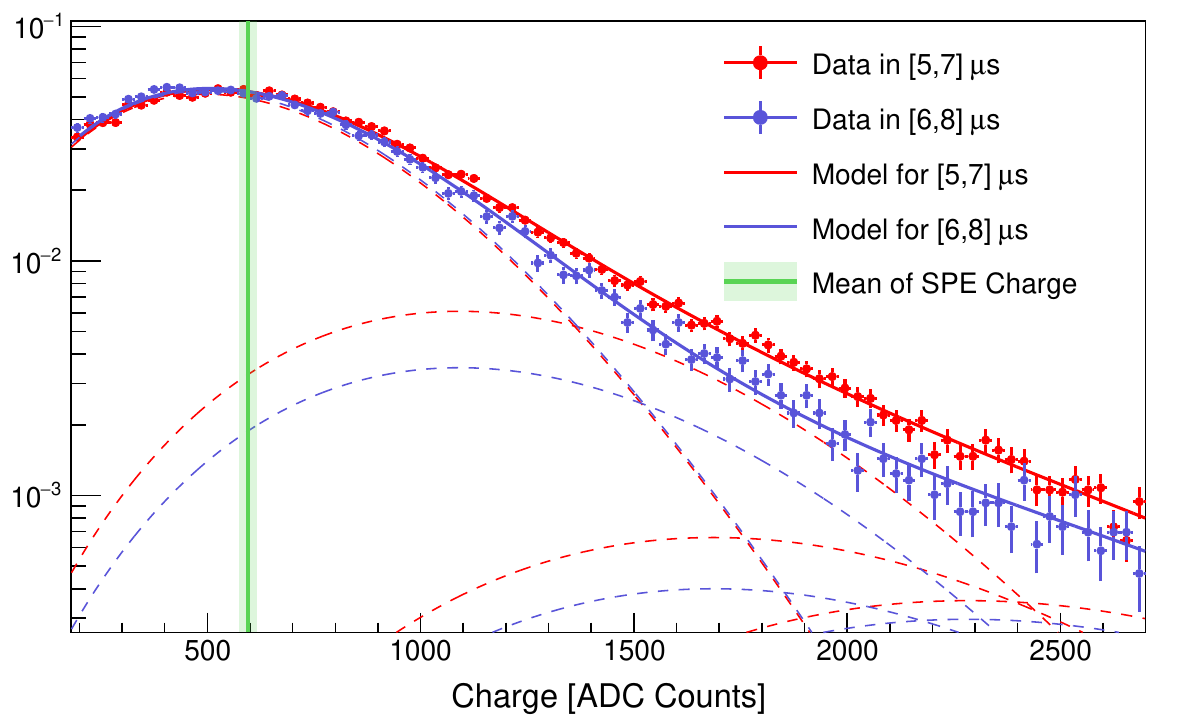}
    \caption{
    {\bf Estimation of mean charge of single photoelectrons (SPEs).} The charge distribution of the isolated clusters in the 5--7\,$\mu$s (red dots) and 6--8\,$\mu$s (blue dots) windows are modeled with up to four NPE clusters, simultaneously fitted for both the 5--7 and 6--8\,$\mu$s windows, yielding a mean charge of 597.2\,ADC for SPE. 
    }
    \label{fig:SPE}
\end{figure}

We measured the light yield of the newly encapsulated COSINE-100U crystal by irradiating it with 59.54\,keV $\gamma$-rays from a $^{241}$Am source. The $^{241}$Am source was positioned above the crystal in Fig.~\ref{fig:testbench}(b). 
The mean charge corresponding to a single photoelectron (SPE) was determined by analyzing trailing isolated cluster pulses in two specific time windows (5--7 and 6--8\,$\mu$s) within 8\,$\mu$s long waveforms. These windows, located 2.6--4.6 and 3.6--5.6\,$\mu$s from the triggering position, were chosen to minimize the impact from large photoelectron clusters.  We simultaneously fit the 5--7 and 6--8\,$\mu$s clusters using models that included up to four photoelectrons clusters~\cite{Choi:2024ziz,Choi:2024trx}, as shown in Fig.~\ref{fig:SPE}. 
The number of photoelectrons (NPE) was calculated by dividing the integrated charge of the main pulse (within 5\,$\mu$s of the pulse start) by the charge of SPE. 

\begin{figure*}[tb!]
    \centering
    \subfigure[]{\includegraphics[width=0.49\textwidth]{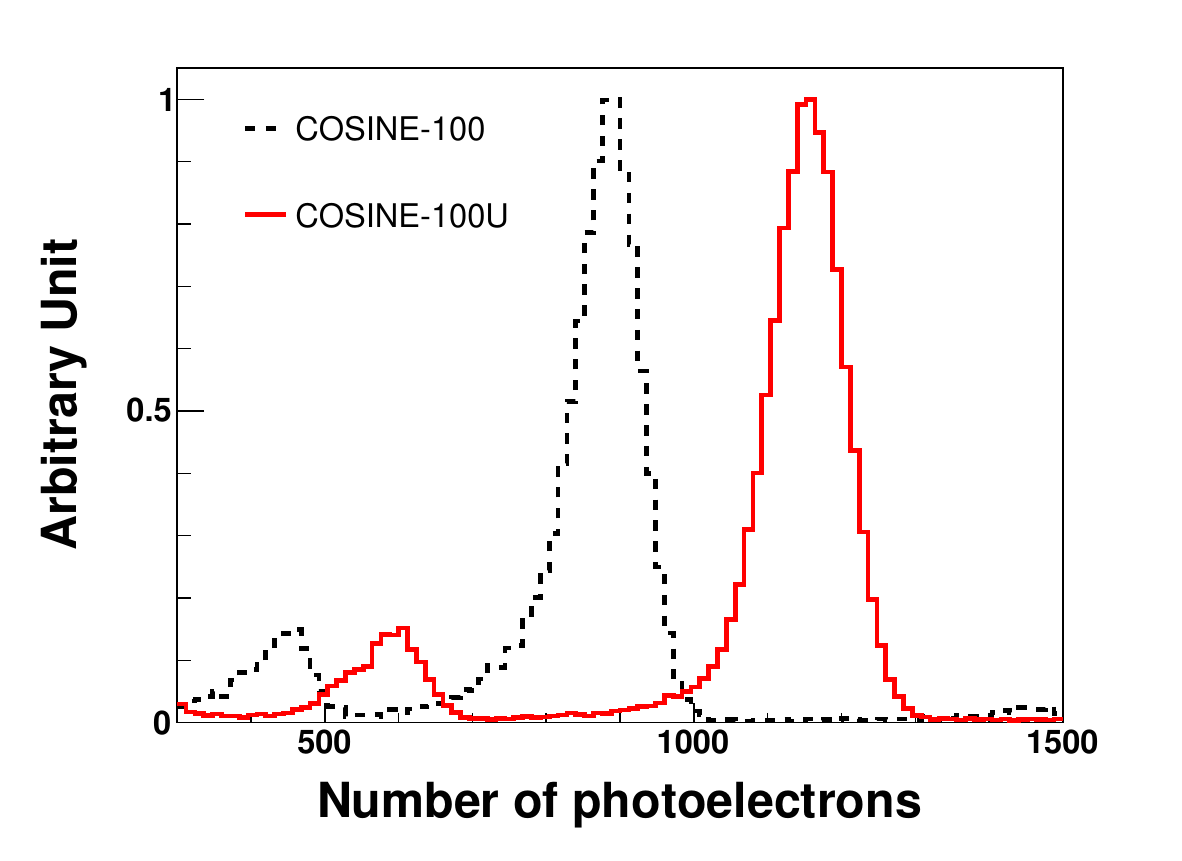}}
    \subfigure[]{\includegraphics[width=0.49\textwidth]{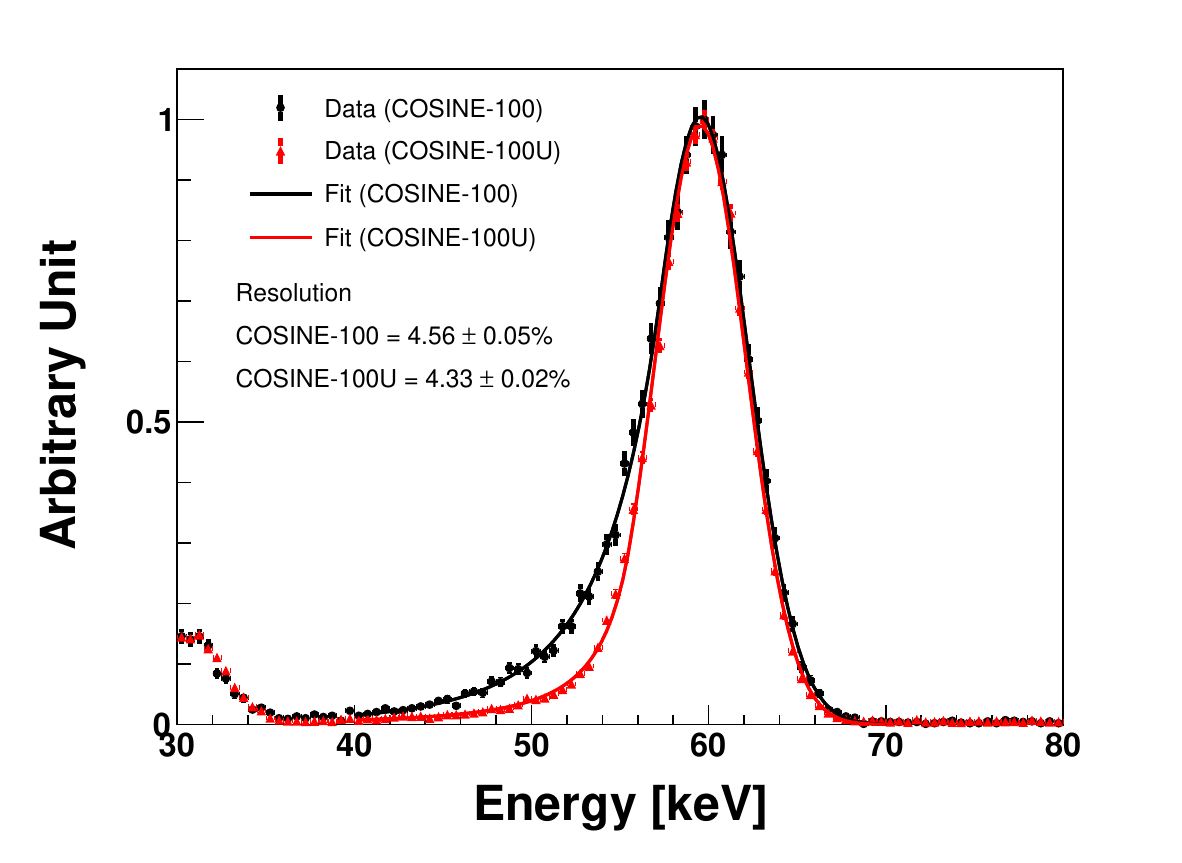}}
    \caption{
    {\bf Light yield improvement of the COSINE-100U encapsulation.} Measured spectra from the C6 crystal in the original COSINE-100 encapsulation and the newly designed COSINE-100U encapsulation at 59.54\,keV using a $^{241}$Am source are compared. (a) Comparison of light yield between COSINE-100 (black dashed line) and COSINE-100U (red solid line). (b) Comparison of energy resolution at 59.54\,keV between the COSINE-100 (black solid line) and COSINE-100U (red solid line) setups. The data have been modeled using Crystal Ball functions to account low energy shoulder caused by partial energy deposition in the encapsulation material.}
    \label{fig:resolution}
\end{figure*}

Figure~\ref{fig:resolution}(a) shows the $^{241}$Am calibration spectra in terms of NPEs, comparing the COSINE-100 setup (black dashed line) and the COSINE-100U encapsulation (red solid line). 
An average increase in light yield of approximately 35$\%$ was observed. Detailed results for each crystal are provided in Table~\ref{table:LY}.

Figure~\ref{fig:resolution}(b) shows the $^{241}$Am calibration spectra in terms of energy, with fit results using a Crystal Ball function to account for the asymmetric shape of 59.54\,keV peak. The asymmetric shape of the lower-energy shoulder is attributed to partial energy deposition in the surrounding encapsulation material due to Compton scattering. For the COSINE-100U C6 crystal, the improved light yield resulted in an approximately 6\% better energy resolution compared to the COSINE-100 setup. 
Similar improvements were observed for all crystals, as summarized in Table~\ref{table:LY}.

In the COSINE-100 experiment, C5 and C8 recorded relatively low light yields due to their 5-inch diameter optical windows, which were initially designed to accommodate 5-inch PMTs. This reduced light yield led to the exclusion of C5 and C8 from the low-energy dark matter search analysis. Additionally, C1 was excluded due to unexpected large noise events from its PMTs. Although the COSINE-100 experiment included eight crystals with a total mass of 106\,kg, the effective detector mass used for the main physics analysis was only 61.4\,kg~\cite{COSINE-100:2021xqn,COSINE-100:2021zqh}.

Upon disassembling C5 and C8 for the COSINE-100U encapsulation, we observed liquid scintillator leakage inside the crystals through thin Mylar windows, which were initially used for low-energy x-ray calibration, such as the 5.9\,keV signal from $^{55}$Fe. This leakage created grooves a few millimeters deep in the crystal surfaces. During the polishing process, we smoothed these grooves, but some remained, resulting in a reduced light yield of approximately 16.5\,NPE/keV for C5 and C8. While this yield is lower than that of other COSINE-100U crystals, it is higher than the light yields of COSINE-100’s good-quality detectors, as summarized in Table~\ref{table:LY}. As a result, these two crystals can now be used for the physics analysis in the COSINE-100U experiment. In addition, C1 was recovered by replacement its PMT, which successfully eliminated the noise issue, thereby increasing the effective detector mass to 99.1\,kg.

\begin{table*}
\centering
\caption{
{\bf Summary of the COSINE-100U energy resolution and light yields.} The root-mean-square (RMS) energy resolution at the 59.54 keV $\gamma$ peak, determined using a $^{241}$Am source calibration, and the measured light yields of the crystals are compared between the COSINE-100 and COSINE-100U setups. The COSINE-100U setup demonstrates a significantly improved energy resolution, attributed to the increased light yield achieved through the upgraded design and enhancements in light collection efficiency. }
\begin{tabular}{ c c  c  c  c  c}
\hline
\multirow{2}{*}{Crystal}   & \multicolumn{2}{c}{Light yield [NPE/keV]} & \multicolumn{2}{c}{59.54\,keV RMS resolution [\%]} \\
     &  COSINE-100  &  COSINE-100U  &  COSINE-100  &  COSINE-100U \\\hline\hline
C1  & 14.9 $\pm$ 1.5 & 22.4 $\pm$ 0.5 &  4.80$\pm$0.06  & 4.41$\pm$0.04 \\ 
C2  & 14.6 $\pm$ 1.5 & 20.1 $\pm$ 0.5 &  5.06$\pm$0.06  & 4.54$\pm$0.04  \\ 
C3  & 15.5 $\pm$ 1.6 & 20.4 $\pm$ 0.4 &  4.96$\pm$0.05  & 4.42$\pm$0.04 \\ 
C4  & 14.9 $\pm$ 1.5 & 20.7 $\pm$ 0.4 &  4.80$\pm$0.05 &  4.59$\pm$0.03\\ 
C5  & 7.3  $\pm$ 0.7 & 16.8 $\pm$ 0.5 &  8.33$\pm$0.11 &  5.04$\pm$0.03 \\ 
C6  & 14.6 $\pm$ 1.5 & 19.6 $\pm$ 0.3 &  4.56$\pm$0.05&  4.33$\pm$0.02\\ 
C7  & 14.0 $\pm$ 1.4 & 20.2 $\pm$ 0.5 &  4.86$\pm$0.04 &  4.53$\pm$0.03\\ 
C8  & 3.5  $\pm$ 0.3 & 16.2 $\pm$ 0.4 & 11.53$\pm$0.11 &  5.06$\pm$0.04\\ \hline
\end{tabular}
\label{table:LY}
\end{table*}

\subsection{Internal Background and Stability Measurement}
\label{section:internal}
Following the $^{241}$Am measurements, we conducted approximately two weeks of background measurements for each crystal in the sea-level shield setup as shown in Fig.~\ref{fig:testbench}. For this, the $^{241}$Am source was removed, as shown in Fig.~\ref{fig:testbench}(b).  Due to the relatively thin layers of lead and liquid scintillator, as well as the high muon flux at sea level, we could not achieve the low-background levels of the COSINE-100 experiment at Y2L. However, we were still able to study internal $\alpha$ background and assess the stability of the encapsulation. 

\begin{figure}[tb!]
    \centering
    \includegraphics[width=1.0\columnwidth]{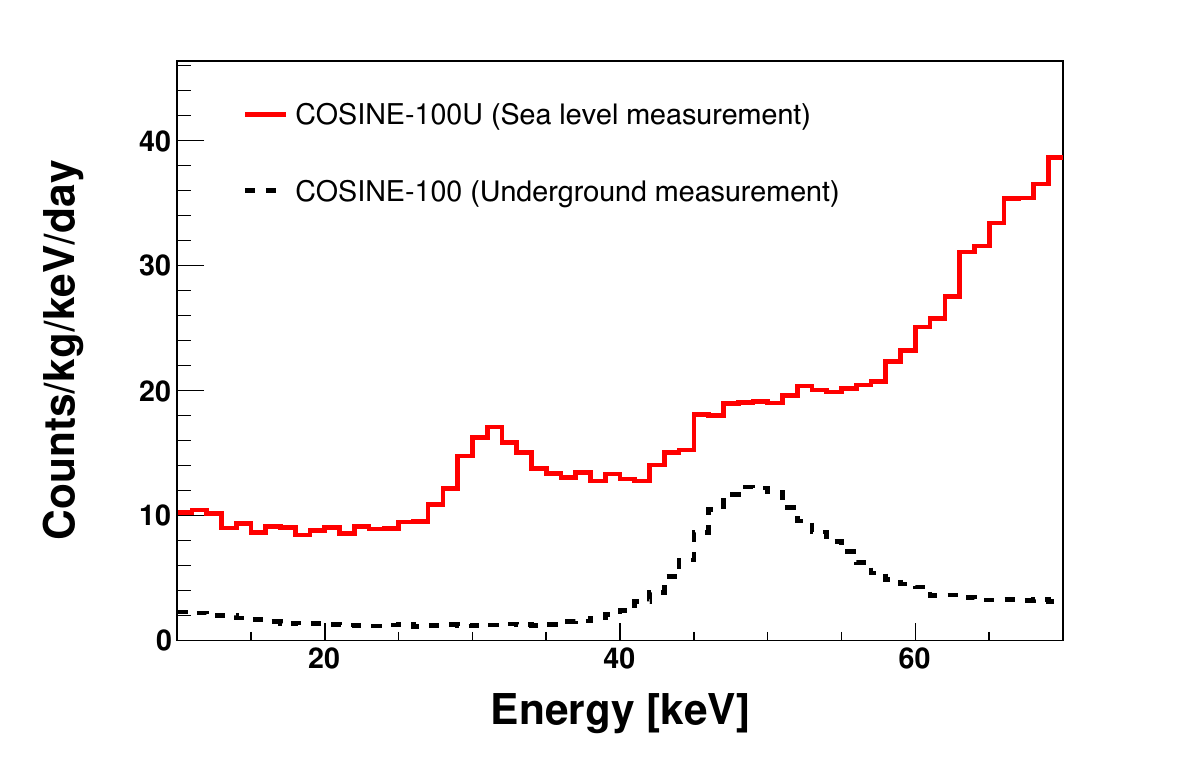}
    \caption{
    {\bf Background measurement at sea level test facility.}
    The single-hit low-energy spectra of C6 from the COSINE-100 experiment (black dashed line) and the sea-level simple shield using the COSINE-100U encapsulation (red solid line) are compared. The distinct 33\,keV peak arises due to cosmogenic activation of $^{121m}$Te and the removal of the quartz window in the new encapsulation, as described in Section \ref{section:internal}. Meanwhile, the 47\,keV peak, originating from internal $^{210}$Pb, remains unchanged.}
    \label{fig:anodesing}
\end{figure}

Events coincident with the liquid scintillator with energies above 80\,keV and within a 200\,ns coincidence window were categorized as multiple-hit events, while all other events were categorized as single-hit events. Figure~\ref{fig:anodesing} shows the single-hit low-energy spectra of C6 from this measurement, using the upgraded COSINE-100U encapsulation (red solid line) compared to the same crystal in the COSINE-100 experiment. As seen in the figure, the sea level measurement had significantly higher background rates due to reduced shielding and increased muon-related backgrounds. A peak around 33\,keV was observed in this sea level measurement with the COSINE-100U encapsulation. This could be due to cosmogenic activation of $^{121m}$Te~\cite{COSINE-100:2019rvp} and external contributions, with the K-shell dip of non-proportional scintillation light~\cite{COSINE-100:2024log} possibly contributing to the peak. Additionally, x-rays from Ba in the PMT glass, as well as Cs and In in the photocathodes, may also contribute. The removal of the 12\,mm thick quartz layer between the crystal and PMTs in the new encapsulation may have enhanced these x-ray signals.

\begin{figure*}[tb!]
    \centering
    \subfigure[]{\includegraphics[width=0.49\textwidth]{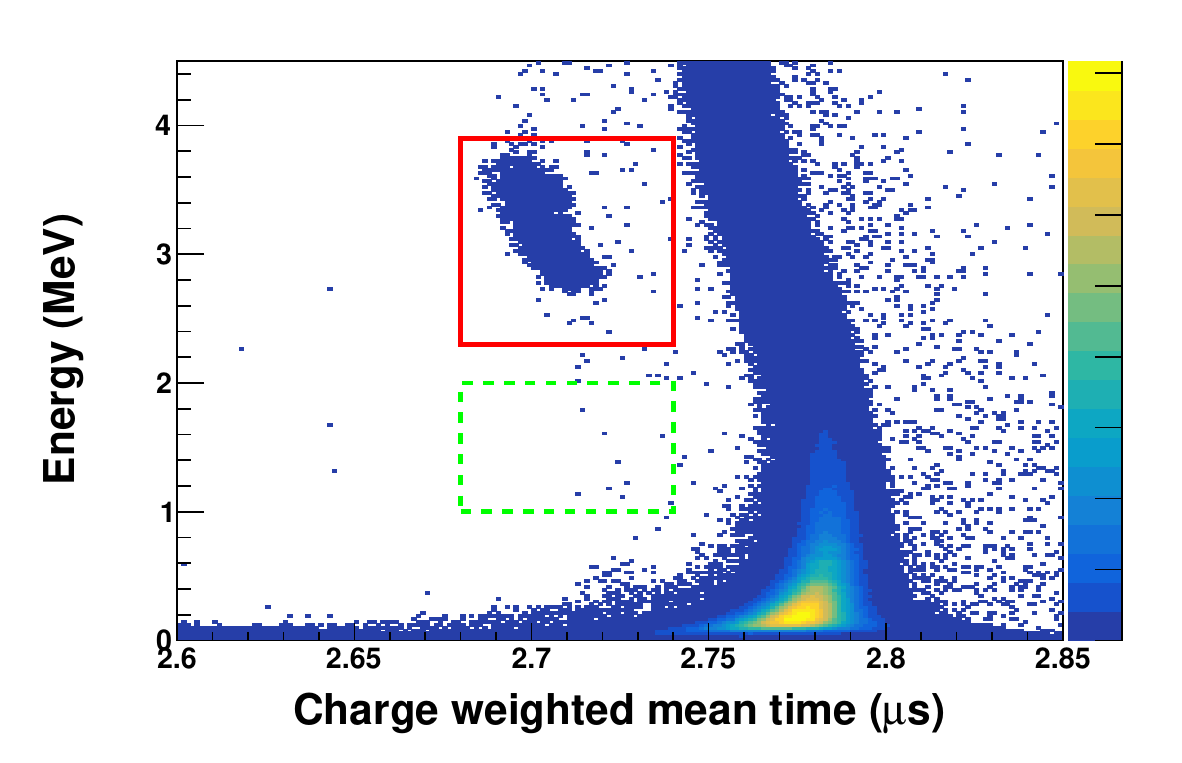}}
    \subfigure[]{\includegraphics[width=0.49\textwidth]{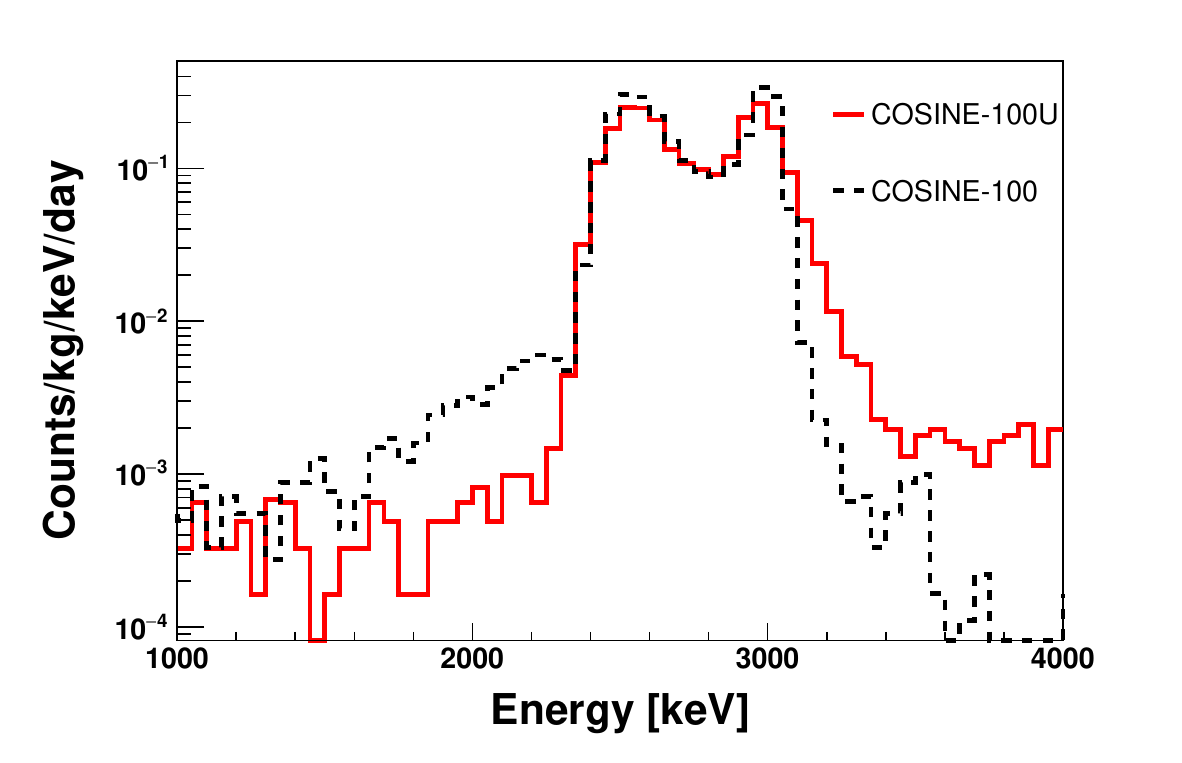}}

    \caption{
    {\bf $\alpha$ measurements.}
    (a) Charge weighted mean decay time is shown for C6. Bulk $\alpha$ (red solid rectangle) and surface $\alpha$ (blue dashed rectangle) are indicated. (b) Energy spectra of $\alpha$ candidate events in this measurement (red solid line) is compared with COSINE-100 setup (black dashed line). The two peaks are caused by $^{210}$Po decays, showing two different $\alpha$ quenching effects observed in the COSINE-100 experiment~\cite{COSINE-100:2023dsf}. }
    \label{fig:alpha}
\end{figure*}

To measure internal $\alpha$ activity, we utilized charge-weighted mean decay time to distinguish between $\alpha$ and beta/gamma events, as shown in Fig.~\ref{fig:alpha}(a), where $\alpha$s form a distinct cluster with shorter decay times, clearly separated from the beta/gamma events. The bulk $\alpha$ contamination from $^{210}$Po is highlighted in the red solid box, while low-energy surface $\alpha$ contamination~\cite{COSINE-100:2023dsf}, with energy in the 1--2\,MeV range, is indicated by the green dashed box. 
Bulk $\alpha$ contamination originates from impurities introduced during the crystal growing process and is expected to be consistent with the COSINE-100 measurements, decreasing over time due to the decay of $^{210}$Pb (with a half-life of 22.3\,years). Surface $\alpha$ contamination, on the other hand, may occur on the crystal surface or on the PTFE reflective sheet during the encapsulation process. We used the event rate of 1--2\,MeV $\alpha$-particles as an indicator of surface contamination, as shown in Fig.~\ref{fig:alpha}(b). Our careful encapsulation process has minimized surface contamination.
Table~\ref{table:alpha} summarizes internal $\alpha$ background measurements of the new COSINE-100U encapsulation compared to the COSINE-100 measurements near shutdown in March 2023. The bulk $\alpha$ measurements show a clear decrease in the COSINE-100U setup, consistent with the decay of internal $^{210}$Pb, while surface $\alpha$ rates are generally lower than in COSINE-100, though some crystals show slightly higher rates. We plan to systematically study surface $\alpha$ contamination by varying surface treatment methods using sample crystals to better understand the causes of contamination.

\begin{table*}
\centering
\caption{{\bf The summary of COSINE-100U $\alpha$ measurements.} Bulk $\alpha$-particles are identified within the energy range of 2.2--3.5\,MeV, whicle surface $\alpha$-particles are selected in the 1--2\,MeV range, as indicated in Fig.~\ref{fig:alpha}.
}
\begin{tabular}{ c  c  c  c  c}
\hline
\multirow{2}{*}{Crystal} & \multicolumn{2}{c}{Bulk alpha [mBq/kg]} & \multicolumn{2}{c}{surface alpha [nBq/cm$^2$]}\\
      &  COSINE-100  &  COSINE-100U  &  COSINE-100  &  COSINE-100U \\\hline\hline
C1 &  2.59 $\pm$ 0.01  &  2.49 $\pm$ 0.02  &  33.07 $\pm$ 3.70  &  45.22 $\pm$ 5.99  \\ 
C2 &  1.69 $\pm$ 0.01  &  1.63 $\pm$ 0.02  &  39.61 $\pm$ 3.76  &  30.42 $\pm$ 10.14 \\ 
C3 &  0.63 $\pm$ 0.01  &  0.59 $\pm$ 0.01  &  71.73 $\pm$ 5.06  &  30.89 $\pm$ 8.57  \\ 
C4 &  0.64 $\pm$ 0.01  &  0.60 $\pm$ 0.01  &  24.27 $\pm$ 2.31  &  51.48 $\pm$ 5.76  \\ 
C5 &          -        &  1.69 $\pm$ 0.02  &          -         &  99.45 $\pm$ 11.72 \\ 
C6 &  1.52 $\pm$ 0.01  &  1.42 $\pm$ 0.01  &  120.9 $\pm$ 5.90  &  39.11 $\pm$ 5.83  \\ 
C7 &  1.51 $\pm$ 0.01  &  1.43 $\pm$ 0.01  &  95.35 $\pm$ 5.24  &  61.70 $\pm$ 7.32  \\ 
C8 &          -        &  1.50 $\pm$ 0.02  &          -         &  24.71 $\pm$ 4.75  \\ \hline
\end{tabular}
\label{table:alpha}
\end{table*}

\begin{figure*}[tb!]
    \centering
    \subfigure[]{\includegraphics[width=0.49\textwidth]{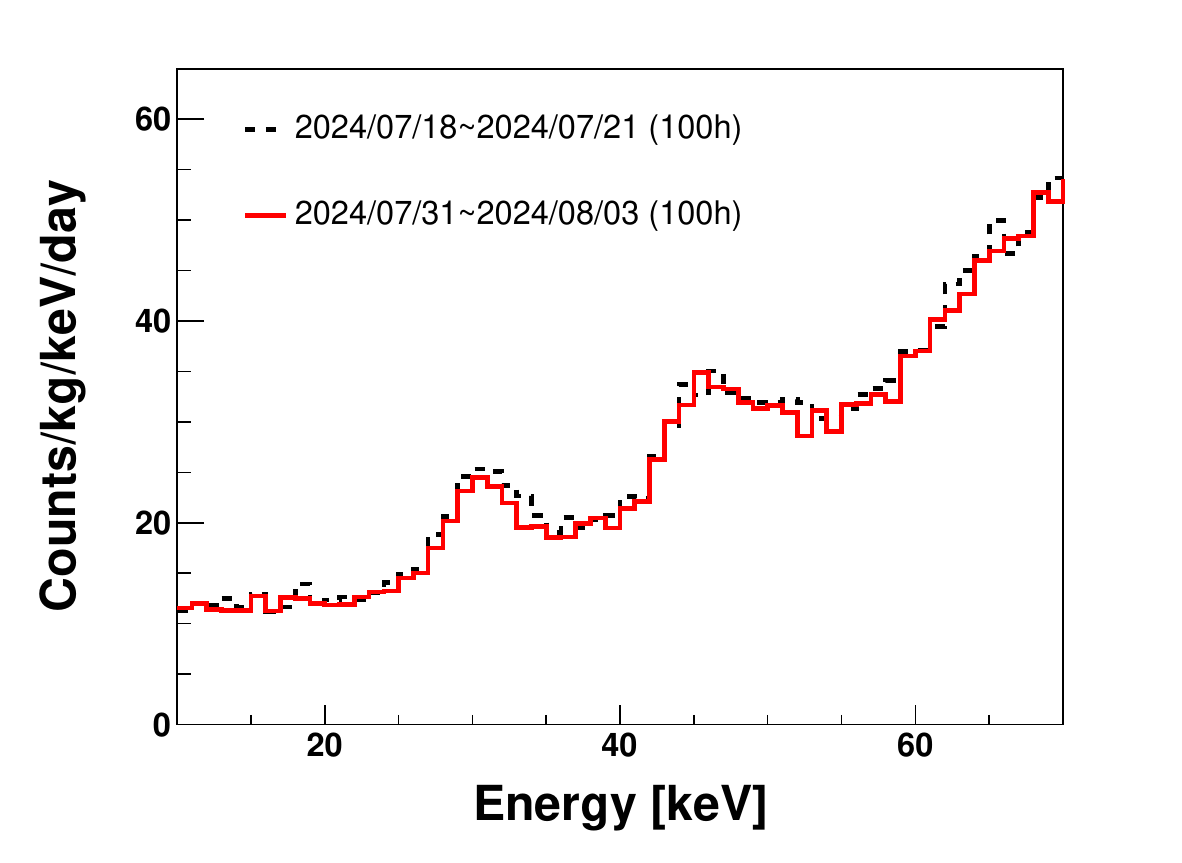}}
    \subfigure[]{\includegraphics[width=0.49\textwidth]{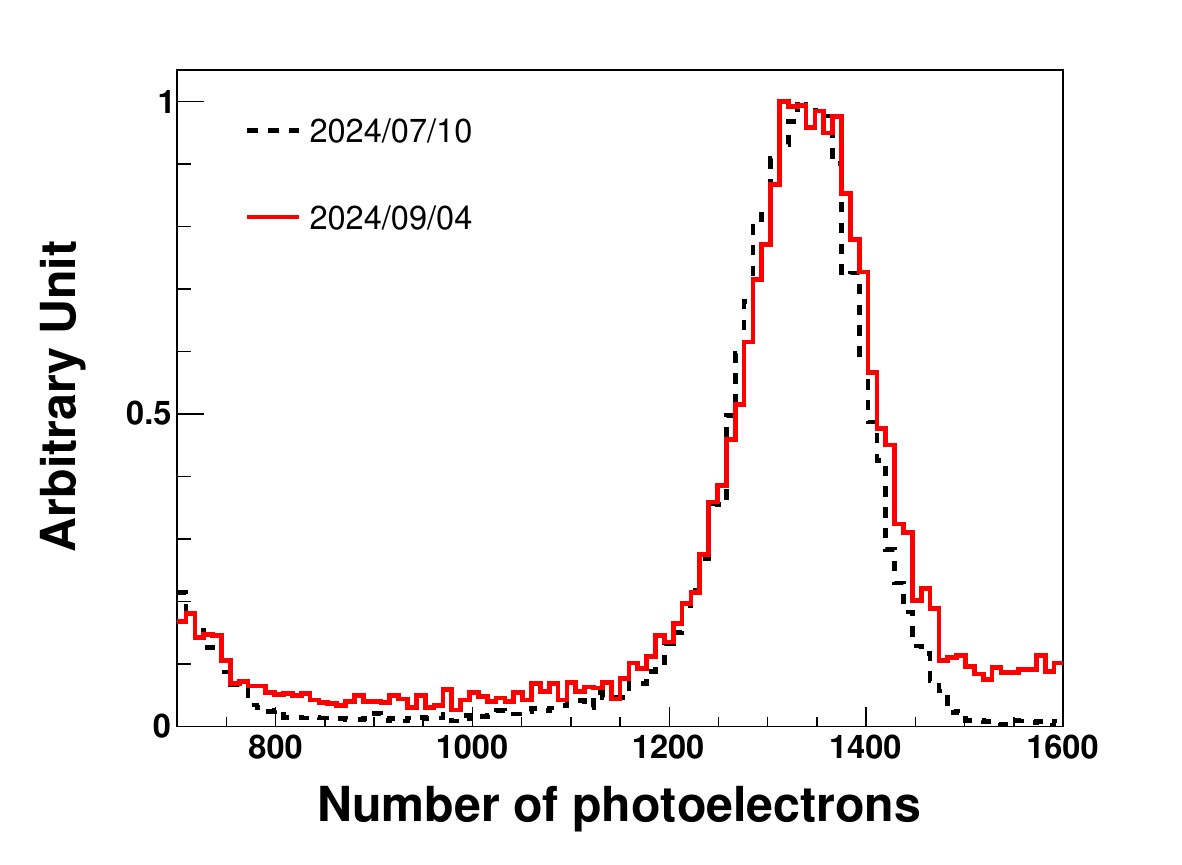}}
    \caption{{\bf Stability of the COSINE-100U encapsulation.} 
    (a) Gain stability during the sea-level background measurements over a two-week period, showing stable conditions with two internal peaks for C1.
(b) The 59.54 keV peak from the $^{241}$Am source measured at both sea level (black dashed line) and Yemilab (red solid line) for C1. Over a two-month period, no significant gain changes were observed in either environment.} 
    \label{fig:stability}
\end{figure*}

The stability of the assembled crystals is monitored by tracking the radiation peaks: 33\,keV, as explained above, and 46.5\,keV from internal $^{210}$Pb.
Figure~\ref{fig:stability}(a) shows the data collected over this period, plotted in 100-hour intervals, demonstrating no noticeable shifts in the peak positions. This indicates that the crystal-PMT coupling remained robust and that no infiltration of liquid scintillator or air occurred.

After completing the background measurements, the upgraded crystals were delivered to Yemilab to minimize cosmogenic activation. The crystals were stored in nitrogen-flushed clean storage. Only the $^{241}$Am source measurement, conducted inside a dark box, was used to monitor any variation in light yield, as shown in Fig.~\ref{fig:stability}(b). We observed consistent light yields from 59.54\,keV peak, indicating stable conditions of the crystal encapsulation.

\section{Yemilab Preparation}\label{sec4}

\subsection{Decommissioning of COSINE-100}
\begin{figure*}[!htb]
  \begin{center}
    \includegraphics[width=1.0\textwidth]{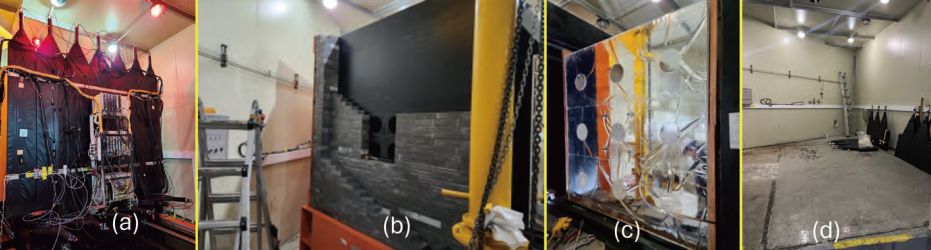}
  \end{center}
  \caption{
        {\bf Decommissioning of the COSINE-100.} Photos were taken during the decommissioning of the COSINE-100 detector at Y2L. (a) The external shield of the COSINE-100 before decommissioning. (b) Removal of outside electronics and muon detectors, with the decommissioning of lead bricks in progress. (c) All external passive shields were removed, and the PMTs for the liquid scintillator active veto detector were detached. (d) Complete removal of the COSINE-100 shield from Y2L. 
    }
\label{fig:decommission}
\end{figure*}

The COSINE-100 experiment, which operated at Y2L, concluded in March 2023 in preparation for the relocation of the experimental site to Yemilab~\cite{Park:2024sio,Yemilab2024} and the detector upgrade for the COSINE-100U experiment. The decommissioning of the detector was completed by October 2023, as shown in Fig.~\ref{fig:decommission}, and all materials were delivered to Yemilab for the installation of COSINE-100U.

\subsection{Yemilab Preparation}
\begin{figure*}[tb!]
    \centering
    \subfigure[\label{fig:yemilaboverview}]{\includegraphics[width=0.49\textwidth]{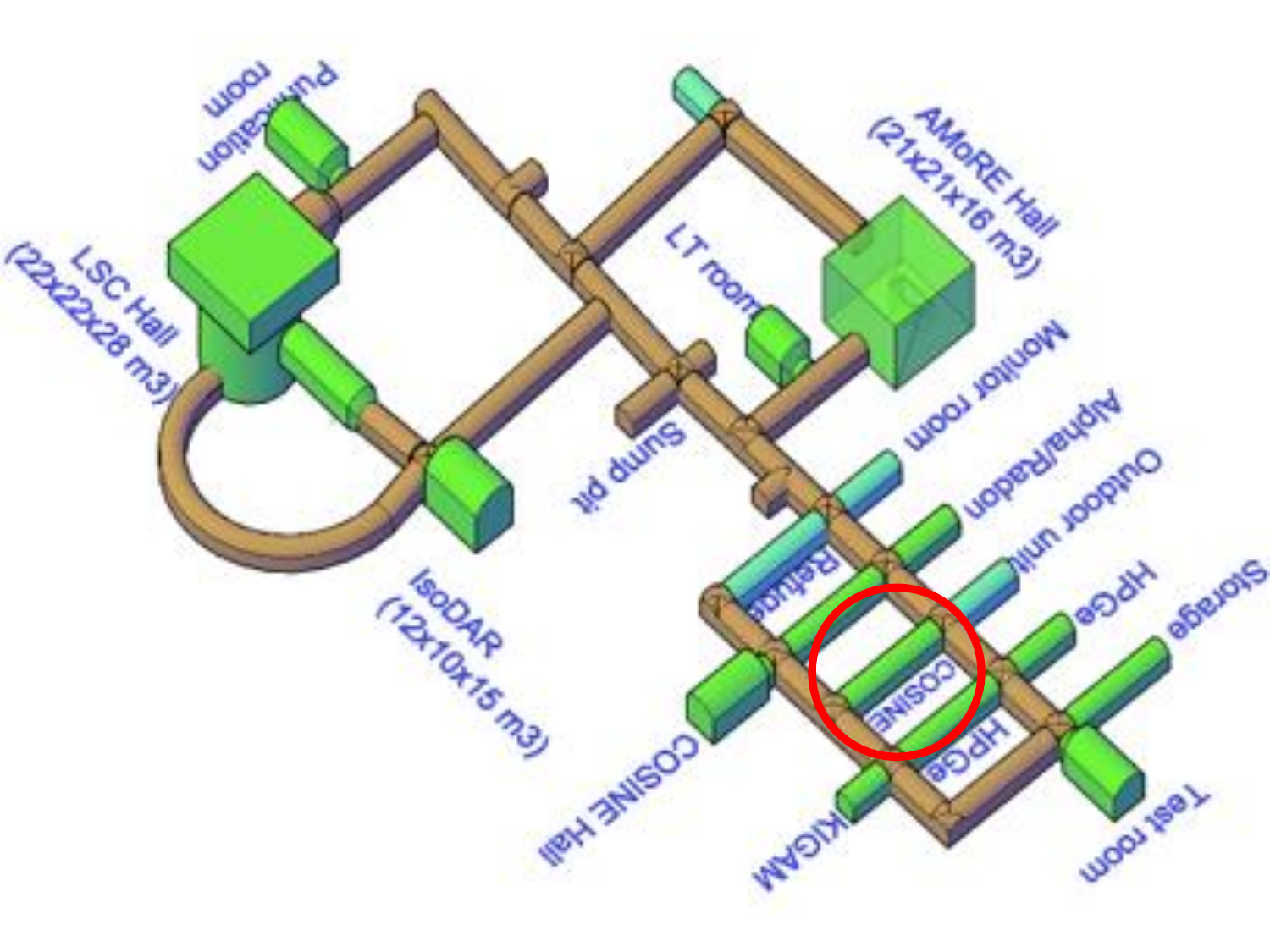}}
    \subfigure[\label{fig:fridgeroom}]{ \includegraphics[width=0.49\textwidth]{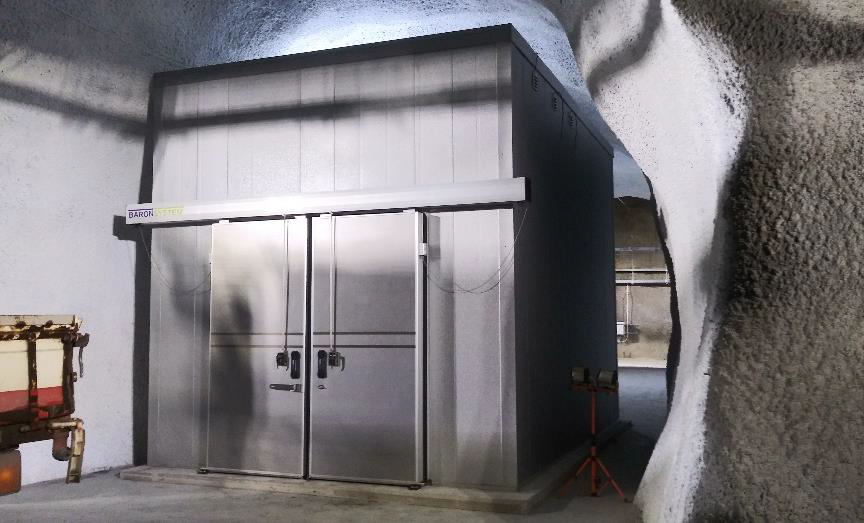}}
    \caption{ {\bf Yemilab for the COSINE-100U. }
    (a) Schematic view of the underground experimental area at Yemilab, highlighting the COSINE-100 tunnel (red circle). (b) The COSINE-100U fridge room, designed to operate at $-$30\,$^{\circ}$C.}
    \label{fig:yemilab}
\end{figure*}

Yemilab is a newly constructed underground laboratory in Korea, completed in September 2022, located in Jeongseon, Gangwon Province, at a depth of 1,000\,m corresponding to 2,700\,m water equivalent~\cite{Park:2024sio,Yemilab2024}.  The facility offers approximately 3,000\,m$^2$ of dedicated experimental space. The underground tunnel accommodates 17 independent experimental spaces, one of which is dedicated to the COSINE-100U experiment, as shown in Fig.~\ref{fig:yemilab}(a). The tunnel can be accessed via a human-riding elevator through a 600\,m vertical shaft and then by electric car through a 780\,m horizontal access tunnel with a 12\% downward slope.  
The surrounding rock is primarily limestone. Ongoing radioactivity measurements of rock samples using inductively coupled plasma mass spectrometry (ICP-MS) and high purity germanium (HPGe) detectors show that the preliminary results are generally consistent with, or slightly lower than, those from Y2L.

The ventilation system at Yemilab efficiently maintains radon levels below 50\,Bq/m$^3$ outside of the summer season. A newly installed radon-reduced air supply system keeps radon levels below 150\,Bq/m$^{3}$ during the summer. 
The post-epoxy floor coating and air filtration system have reduced PM10 dust levels to below 10\,$\mu$g/m$^3$, well within typical office environmental standards. Stricter controls aim to further reduce dust level to below 5\,$\mu$g/m$^3$. Additionally, Yemilab features a Radon Reduction System (RRS) supplying 50\,m$^{3}$/h of air with radon levels below 100\,mBq/m$^3$, which will be used in the COSINE-100U detector room. 

Preliminary measurements of the muon flux in Yemilab indicate a flux of 1.0$\times 10^{-7}\mu$/cm$^2$/s, which is four times lower than the muon flux in Y2L, measured at 3.8$\times 10^{-7}\mu$/cm$^2$/s~\cite{Prihtiadi:2017inr}. Overall, the background environments at Yemilab are significantly better than that of Y2L, leading to reduced external radioactive background contributions.  

We have prepared a warehouse-type refrigerator with a 10\,kW cryocooler to serve as the COSINE-100U detector room, as shown in Fig. ~\ref{fig:yemilab}(b). The plan is to operate the COSINE-100U experiment at $-$30\,$^{\circ}$C to enhance light yield and improve pulse shape discrimination for nuclear recoil events~\cite{Lee:2021aoi}. Based on previous measurements at  $-$35\,$^{\circ}$C, we expect an increase in light yield of approximately 5\% compared to operation at room temperature for electron recoil events. Additionally, an increase in the $\alpha$ quenching factor of approximately 9\% was observed, suggesting a potential further improvement in light yield for nuclear recoil events. While cooling to $-$35\,$^{\circ}$C is technically feasible, operating at $-$30\,$^{\circ}$C was choosen to avoid excessive load on the cryocooler. The COSINE-100U detector room measures 4\,m in width,  6\,m in length, and 4\,m in height, and is located at the front of the COSINE tunnel. 

\subsection{Shielding Installation}
Inside the COSINE-100U fridge room, shielding was installed to protect the experiment from external radiation sources and to provide an active veto for internal or external contamination~\cite{Adhikari:2017esn}. Most of the shielding components from the COSINE-100 experiment were recycled for use in the COSINE-100U setup. This shielding consists of a four-layer nested arrangement, starting from the inside: 40 cm of liquid scintillator, 3 cm of copper, 20 cm of lead, and 3 cm of plastic scintillator. The liquid scintillator~\cite{Adhikari:2020asl} and plastic scintillator~\cite{Prihtiadi:2017inr} layers actively tag radioactivity from internal contamination, external radiation, and muon events.

\begin{figure}[tb!]
    \centering
    \includegraphics[width=1.0\columnwidth]{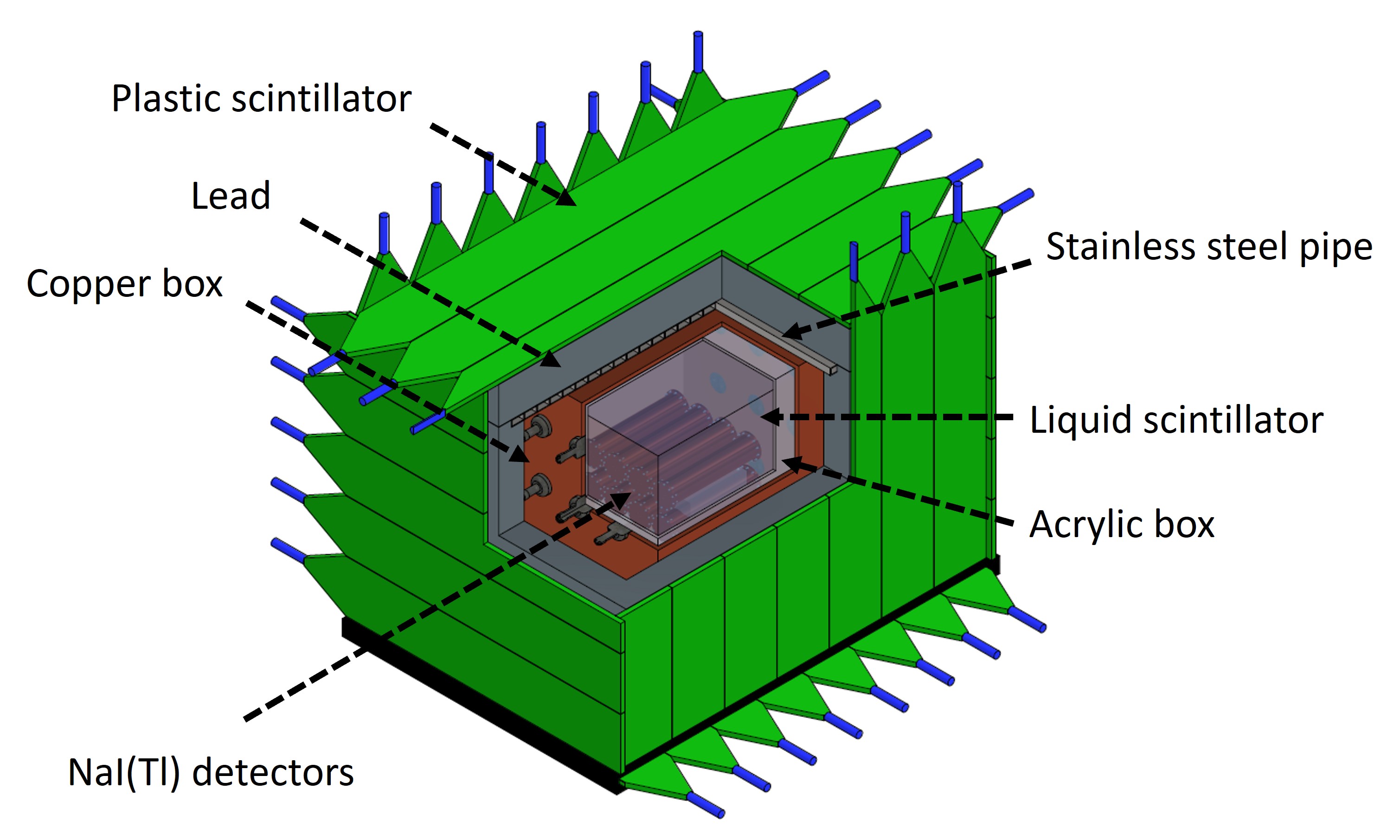}
    \caption{
    {\bf Schematic of the COSINE-100U detector.}
      The setup includes eight NaI(Tl) detectors with a total mass of 99.1\,kg, immersed in 2,200\,L of liquid scintillator. The detectors are surrounded by multiple layers of shielding: 3\,cm of copper, 20\,cm of lead, and 3\,cm of plastic scintillator, effectively minimizing background contributions. }
    \label{fig:COSINE-100U_Schematic}
\end{figure}

\begin{figure}[tb!]
    \centering
    \includegraphics[width=1.0\columnwidth]{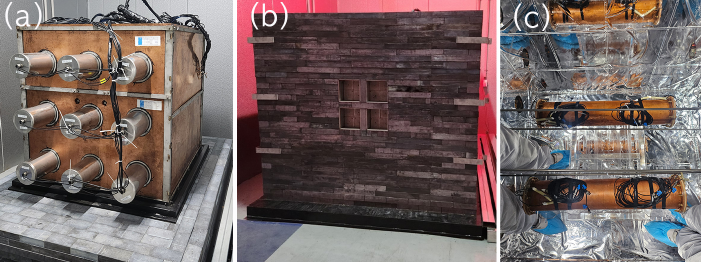}
    \caption{
    {\bf Shield installation of COSINE-100U at Yemilab.} 
    (a) The liquid scintillator active veto detector, composed of a 1\,cm thick acrylic box surrounded by 3\,cm think OFC, was installed with 5-inch PMTs for readout. (b) A 20\,cm thick lead layer surrounds the copper container for the liquid scintillator. (c) Inside of the acrylic box for the liquid scintillator, the C5 and C8 crystals were installed.   }
    \label{fig:COSINE-100U_Shield}
\end{figure}

The COSINE-100 shield utilized a steel skeleton to support heavy elements and allow access to the inner structure with a mechanical opening system~\cite{Adhikari:2017esn}. However, this design inherently included approximately 4\,tons of steel inside the lead shields. In contrast, the COSINE-100U shield does not use a steel skeleton. 
Instead,  heavy materials, such as lead bricks, are stacked on a precisely leveled steel plate, similar to the shield used in the NEON experiment~\cite{NEON:2022hbk}, as shown in Figs.~\ref{fig:COSINE-100U_Schematic} and \ref{fig:COSINE-100U_Shield}. 
To reinforce the top structure, 5\,cm$\times$10\,cm square stainless steel pipes, each 180\,cm in length, will support the lead bricks. 
Figure~\ref{fig:COSINE-100U_Schematic} illustrates the overall detector geometry of the COSINE-100U setup, while Fig.~\ref{fig:COSINE-100U_Shield} shows a photograph of the COSINE-100U shield  during installation at Yemilab.

We produced 2,400\,L Linear Alkyl-Benzene(LAB) based liquid scintillator~\cite{Kim:2024spf}, following a recipe similar to that used in the COSINE-100 experiment~\cite{Adhikari:2017esn,Adhikari:2020asl}.  
The old COSINE-100 liquid scintillator will be repurposed for test measurements facilities at Yemilab. 

\subsection{Physics Operation Plan}
For the operation of COSINE-100U, all electronics--including preamplifiers, FADCs, high-voltage power supplies, and the computer server for data acquisition--will be installed in a $-$30$^{\circ}$C environment. We have tested all these electronics in a $-$30$^{\circ}$C environment for three months, and no issues were observed.

Increased light yield and improved pulse shape discrimination of nuclear recoil events for NaI(Tl) crystals at $-$30$^{\circ}$C were observed, as reported in Ref.~\cite{Lee:2021aoi}. Preliminary tests with a small liquid scintillator cell also showed an increased light yield, consistent with results from the literature~\cite{Xia:2014cca}.

However, initial measurements of two crystals (C5 and C8) installed at Yemilab under $-$30$^{\circ}$C operation, as shown in Fig.~\ref{fig:COSINE-100U_Shield}(c), revealed weakening of the sealing through the PTFE gasket. This issue may be due to differential thermal contraction between the PTFE gasket and the OFC lid. To address this, the PTFE gasket is being replaced with a Viton O-ring to ensure reliable sealing at low temperatures.

We assembled a test crystal from the same manufacturer as the COSINE-100 crystals using the Viton O-ring to assess stability at room temperature. During two months of measurements, no performance issues were observed. The modification of the gaskets was applied to all COSINE-100U crystals.

Once all detectors are assembled, we will proceed with the installation of all COSINE-100U components, including the eight crystals, the liquid scintillator, and the top lead bricks and outer muon plastic scintillator panels. If the schedule proceeds as planned, the COSINE-100U experiment will begin physics operations at room temperature in 2025.

Simultaneously, the test crystal will undergo stability checks at low temperature using a refrigerator installed at IBS. A test period of at least six months is planned to confirm its long-term stability. Once sufficient confidence in low-temperature operation is achieved, the experiment will transit to low-temperature operation.

\section{Expected Background }\label{sec5}
\begin{figure*}[tb!]
    \centering
       \subfigure[]{\includegraphics[width=0.406\textwidth]{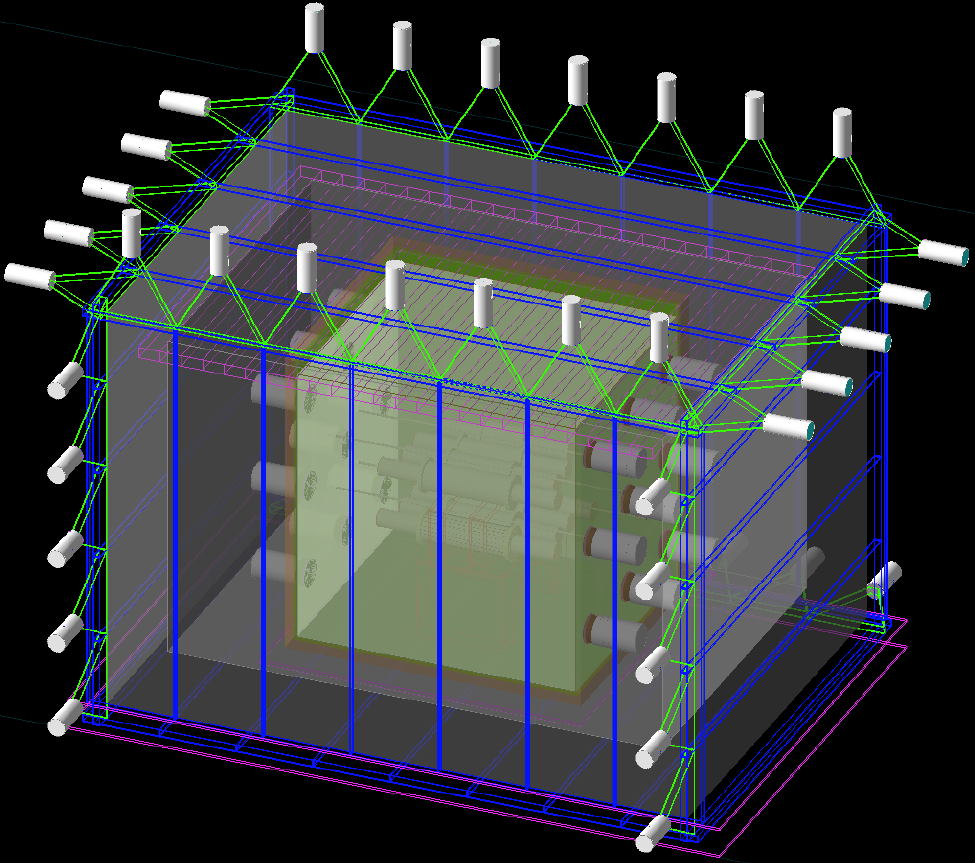}}
    \subfigure[]{\includegraphics[width=0.584\textwidth]{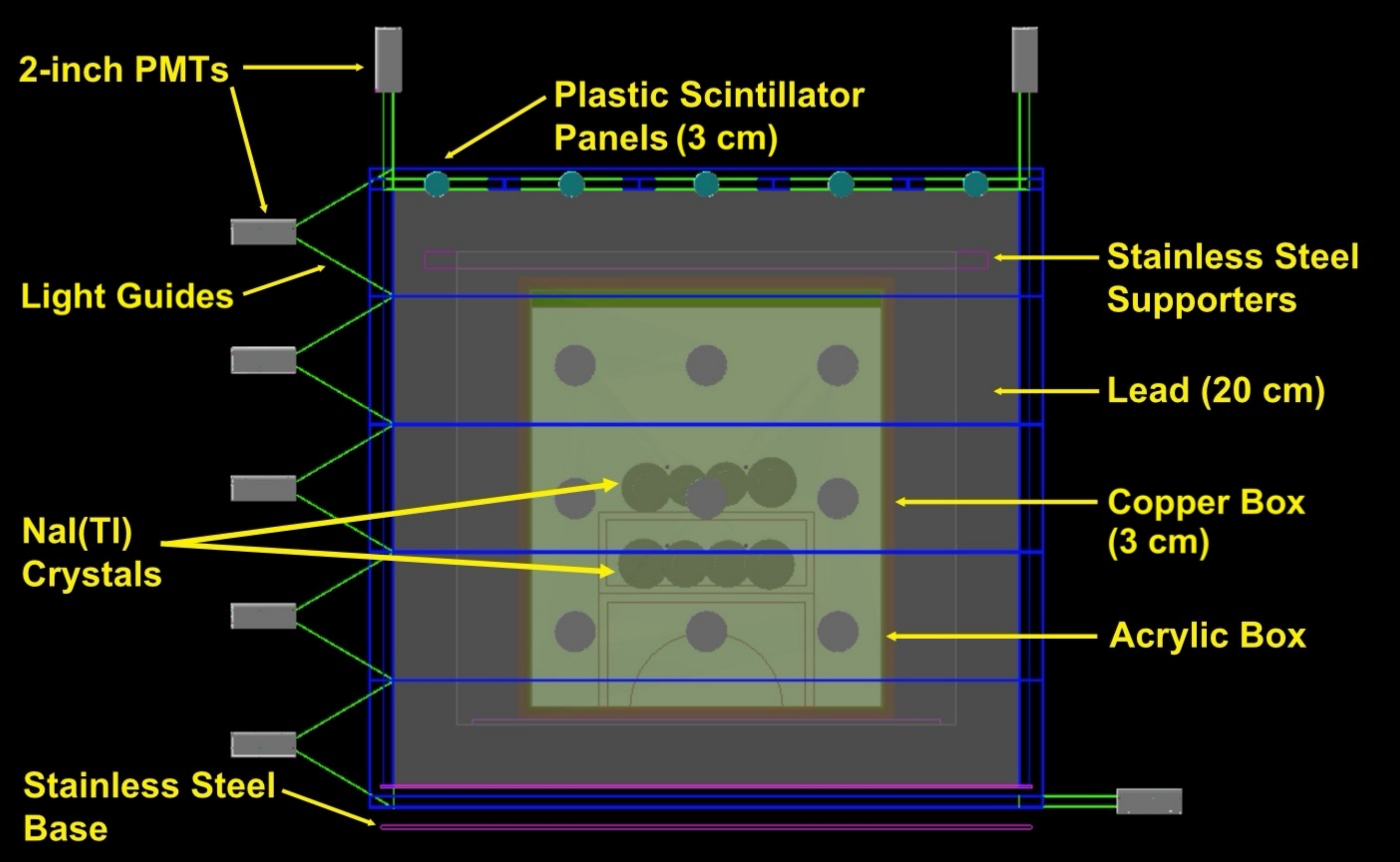}}
    \caption{
     {\bf COSINE-100U detector geometry used in the Geant4 simulation.} (a) Bird's-eye view and (b) front view. Eight NaI(Tl) crystals are supported by the acrylic table inside the liquid scintillator. Nine 5-inch PMTs in each side are attached to read photons from the liquid scintillator. The outer lead shield and plastic scintillators are also installed. To support the shielding on the top, stainless steel support pipes are installed. The entire shielding setup is placed on a precisely labeled steel base. } 
    \label{fig:geant4_geo}
    \end{figure*}

We have gained a precise understanding of the backgrounds in the COSINE-100 detector through Geant4-based simulations~\cite{cosinebg,cosinebg2,COSINE-100:2024ola}. To account for COSINE-100U-specific background contributions, we constructed detector geometry for use in the Geant4-based simulation, as shown in Fig.~\ref{fig:geant4_geo}.  Since the COSINE-100U experiment uses the same crystals as COSINE-100, with only minor machining and surface polishing, we expect the majority of background contributions in COSINE-100U, particularly from internal contaminants, to be very similar to those observed in the COSINE-100 experiment.

However, a few differences are expected due to the redesigned crystal encapsulation. The encapsulation components were replaced, and an additional inner PTFE structure was incorporated. We measured the radioactivity levels of all encapsulation components, as summarized in Table~\ref{table:radioactivity}. As we carefully selected all materials, the contamination levels of the new encapsulation materials are much lower than those of the PMT and PMT base. Based on our understanding of the COSINE-100 backgrounds~\cite{COSINE-100:2024ola} and the measured contamination levels of the encapsulation materials, we simulated the expected background contributions of radiation components. As in COSINE-100, the PMTs remain the dominant source of external background contamination. 

\begin{table}
\centering
\caption{{\bf Measured radioactivity contaminants in detector components inside the shielding.} The radioactivities were measured using high-purity germanium detectors at Y2L assuming secular equilibrium; upper limits are quoted with a 90\% confidence level. The PMTs, PMT bases, and Viton O-ring are measured in units of mBq/unit, while the other components are measured in units of mBq/kg.    }
\begin{tabular}{c c c  c }
\hline 
  \textbf{Material}& $^{238}$U  & $^{232}$Th  & $^{40}$K \\ \hline\hline
  Copper & 0.27 ± 0.07 & 0.25 ± 0.07 & $<$ 1.24 \\
  PTFE &  $<$ 0.19 &$<$ 0.38 & 8.88 ± 1.32  \\
  Brass bolts & $<$ 0.7 & $<$ 0.7 & $<$ 6.1  \\
  Optical pad &$<$ 1.95&$<$ 4.13 & $<$ 33.5 \\
  Quartz & $<$ 1.8 & $<$ 2.0 & $<$ 20.0\\
  PMTs &   60 ± 10 & 12 ± 5 & 58 ± 5  \\
  PMT Base &  12.2 ± 2.5 &1.7 ± 0.4 & 6.2 ± 2.5  \\
  Viton O-ring &  3.5 ± 0.4 &0.8 ± 0.1 & 4.8 ± 1.2  \\

\hline
\end{tabular}
\label{table:radioactivity}
\end{table}

The polishing of all crystal surfaces and the replacement of the Teflon lapping films may result in different surface contamination levels in the COSINE-100U crystals. Generally, we observed fewer $\alpha$ particles with partial energy deposition (1–2\,MeV measured energy), which may suggest lower surface contamination (see Section~\ref{section:internal}). However, for this study, we conservatively assume the same surface contamination background contributions as observed in the COSINE-100 experiment.

\begin{figure*}[tb!]
    \centering
    \subfigure[]{\includegraphics[width=0.49\textwidth]{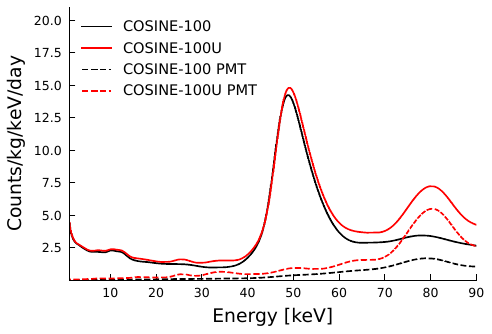}}
    \subfigure[]{\includegraphics[width=0.49\textwidth]{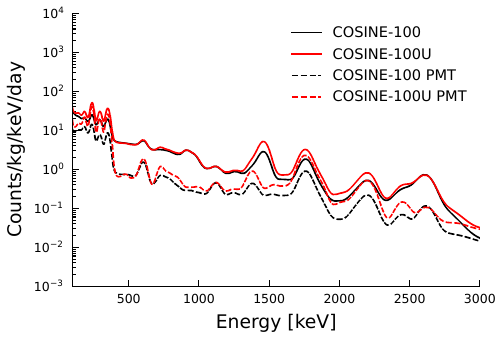}}
    
    \caption{
     {\bf Expected energy spectra of the C6 crystal.} The expected background level of the C6 in the COSINE-100U setup compared with the COSINE-100 background~\cite{COSINE-100:2024ola}. (a) Low-energy spectra from the anode readout of COSINE-100 (black solid line) compared with the expected background of the COSINE-100U setup (red solid line). The main differences in PMT contributions are separated as dashed lines.
 (b) High-energy spectra from the dynode readout are presented in the same manner. Although an increase in background contributions from the PMTs is expected in the COSINE-100U encapsulation, its impact on the signal region below 6\,keV is negligible.   } 
    \label{fig:bkgdexp}
    \end{figure*}

Although the same PMTs are used, the removal of the 12\,mm quartz layer could potentially increase background contributions from the PMTs to the crystals. We simulated these background contributions in the COSINE-100U geometry (Fig.~\ref{fig:geant4_geo}), assuming the same contamination levels as in the COSINE-100 experiment~\cite{COSINE-100:2024ola}. The absence of the 12\,mm quartz shield may enhance the x-ray contribution from the PMTs, but this effect is primarily observed at energies above 20\,keV, with no significant differences in the signal region below 6\,keV, as shown in Fig.~\ref{fig:bkgdexp}.

\section{Sensitivity of the COSINE-100U experiment}\label{sec6}
With the improved performance of higher light yields in the COSINE-100U detectors, along with background levels similar to those observed in the COSINE-100 experiment in the low-energy signal region, we evaluate the sensitivity of the COSINE-100U experiment for detecting dark matter, particularly for spin-dependent WIMP-proton interaction. We assume one year of operation, using the measured light yields at room temperature as summarized in Table~\ref{table:LY}, and the expected background levels discussed in Section~\ref{sec5}, based on the COSINE-100 measurement~\cite{COSINE-100:2024ola}.  

The current COSINE-100 data analysis achieved an 8\,NPE threshold~\cite{Yu:2024qgq}; however, further improvements are expected through the use of machine learning techniques and simulated waveforms of NaI(Tl) crystals~\cite{Choi:2024ziz}. These advancements are expected to lower the analysis threshold to 5\,NPE, a level already achieved by the COHERENT experiment with a CsI(Na) crystal~\cite{COHERENT:2017ipa}. For the sensitivity analysis of the COSINE-100U experiment, we assume a 5\,NPE analysis threshold for each crystal. 
Additionally, we evaluate sensitivities for different energy thresholds, including 8\,NPE for the COSINE-100U setup and 8\,NPE for the COSINE-100 setup, to illustrate the substantial improvements provided by the upgraded configuration. 

\begin{figure*}[tb!]
    \centering
    \subfigure[]{\includegraphics[width=0.49\textwidth]{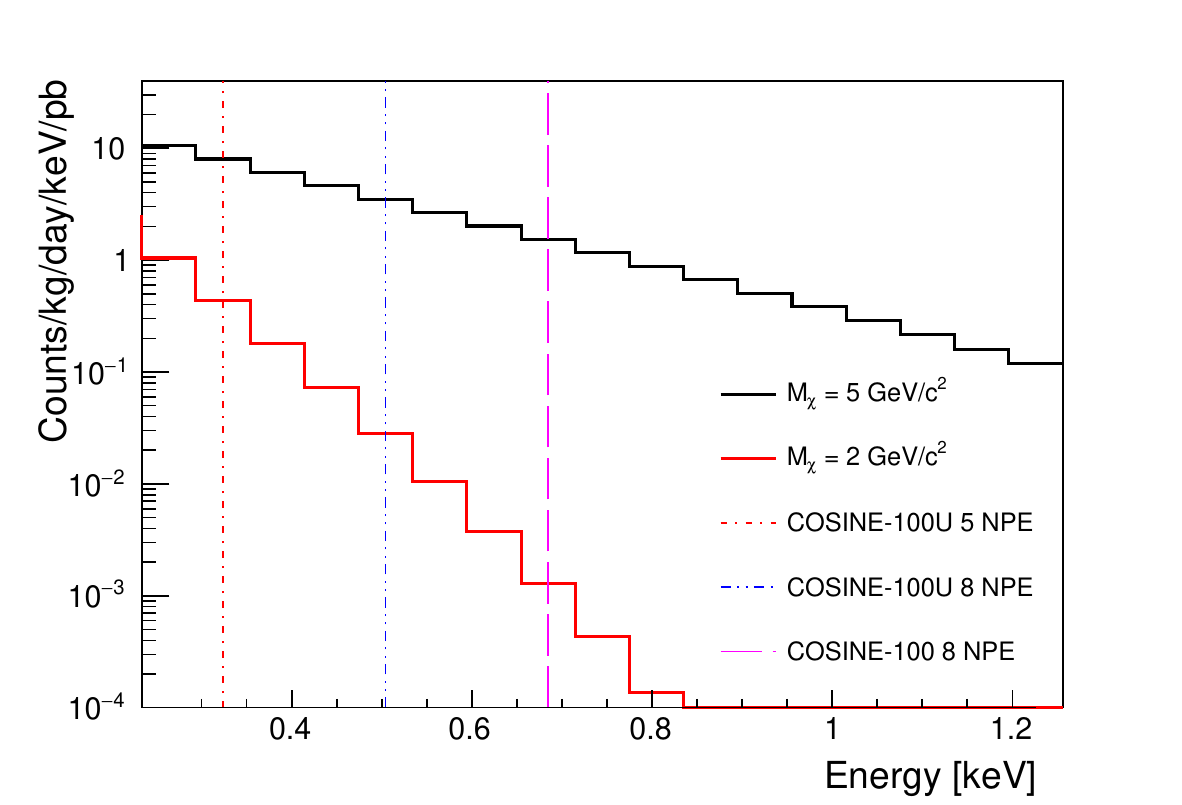}}
    \subfigure[]{\includegraphics[width=0.49\textwidth]{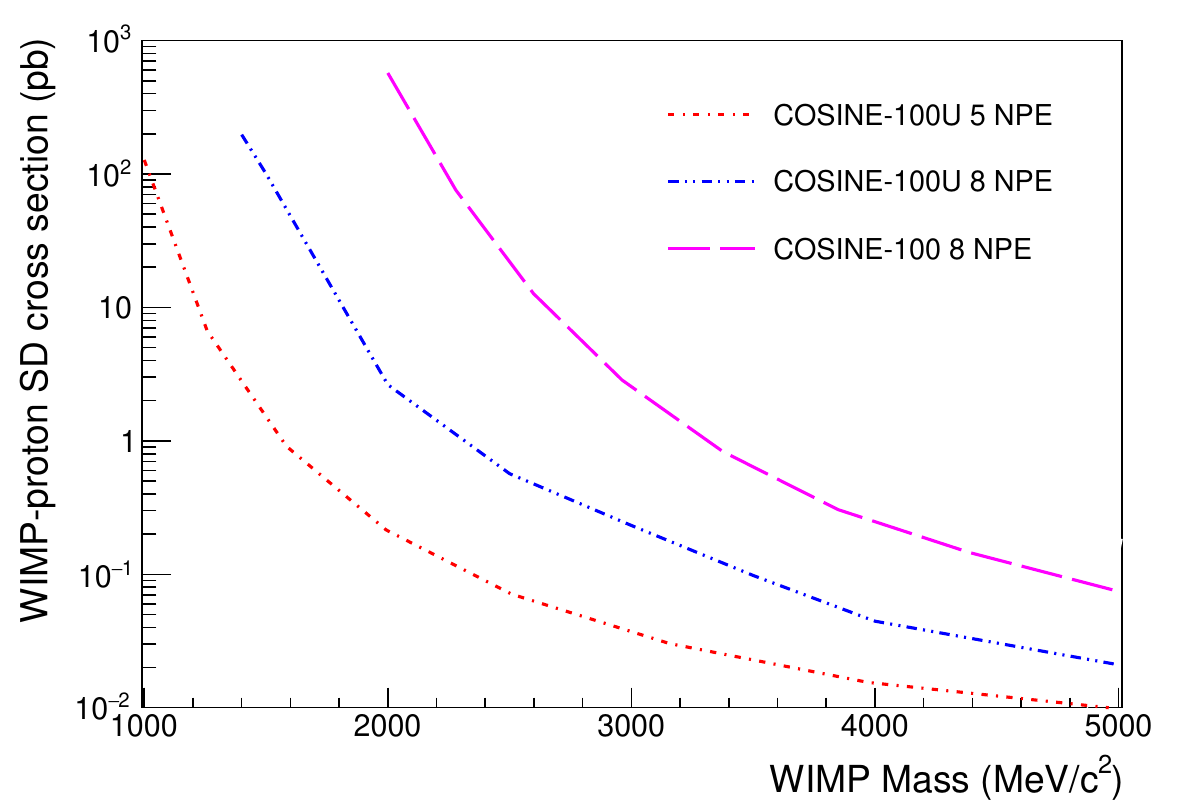}}
    
    \caption{
     {\bf Expected signal rates and sensitivities with different energy thresholds.} (a) The expected signal rates for two WIMP masses (M$_{\chi}$) of 2 and 5\,GeV/c$^2$ are shown for three  energy thresholds: COSINE-100 (8\,NPE), COSINE-100U (8\,NPE), and COSINE-100U (5\,NPE), using the C6 crystal.  
 (b)  The expected 90\% confidence level limits for the WIMP-proton spin-dependent cross-section are shown, assuming a 99.1\,kg NaI(Tl) target operated for one year.  The limits are presented for three different energy thresholds: COSINE-100 with 8\,NPE, COSINE-100U with 8\,NPE, and COSINE-100U with 5\,NPE. } 
    \label{fig:sigexp}
    \end{figure*}

We generate WIMP interaction signals with and without the Midgal effect~\cite{Migdal,Ibe:2017yqa,COSINE-100:2021poy}. These signals are simulated for various interactions and masses within the standard WIMP galactic halo model~\cite{Lewin:1995rx,Freese:2012xd}. Form factors and proton spin values of the nuclei are implemented using the publicly available {\sc dmdd} package~\cite{dmdd,Gluscevic:2015sqa,Anand:2013yka,Fitzpatrick:2012ix,Gresham:2014vja}. The electron-equivalent energy of the nuclear recoil is reduced using nuclear recoil quenching factors, which represent the ratio of scintillation light yield from sodium or iodine recoil relative to that from electron recoil for the same energy. Recently measured quenching factor values~\cite{Lee:2024unz} are used, and  the inclusion of the Migdal effect in the NaI(Tl) crystals follows our previous study~\cite{COSINE-100:2021poy}. Figure~\ref{fig:sigexp}(a) shows the expected signal rates for two benchmark WIMP masses (2 and 5\,GeV/c$^2$) in spin-dependent WIMP-proton scattering scenarios without the Migdal effect. The measured electron-equivalent energy is converted to NPE using the light yield data in Table~\ref{table:LY} and nonproportionality measurement in Ref.~\cite{COSINE-100:2024log}. 

Poisson fluctuations in the measured NPE are considered for detector resolution, using a waveform simulation package~\cite{Choi:2024ziz}, which has been validated with the low-energy signal region of COSINE-100 data~\cite{COSINE-100:2024log}. We use an ensemble of simulated experiments to estimate the sensitivity of the COSINE-100U experiment, expressed as the expected cross-section limits for the WIMP-proton spin-dependent interactions in the absence of signals. For each experiment, a simulated spectrum is generated under a background-only hypothesis based on assumed background levels. Gaussian fluctuations of background components from the COSINE-100 measurement~\cite{COSINE-100:2024ola}, along with COSINE-100U-specific background contributions discussed in Section~\ref{sec5}, and Poisson fluctuations in each energy bin are incorporated into each simulated experiment. 

We then fit the simulated data with a signal-plus-background hypothesis, applying flat priors for the signal and Gaussian constraints for the backgrounds. Systematic uncertainties affecting the background model are included as nuisance parameters~\cite{COSINE-100:2021xqn}.  A Bayesian approach is used to analyze the single-hit energy spectrum between 5\,NPE (or 8\,NPE) and 130\,NPE for each WIMP model, covering several WIMP masses. Marginalization was performed to obtain the posterior probability density function for each simulated sample, allowing us to set the 90\% confidence level exclusion limits. 

To evaluate performance improvements, we consider three scenarios for detector performance, as shown in Fig.~\ref{fig:sigexp}. The improved COSINE-100U encapsulation achieves approximately 80 times better sensitivity for a WIMP mass of 2\,GeV/c$^2$,  assuming the same 8\,NPE threshold as in the COSINE-100 analysis~\cite{Yu:2024qgq}. Lowering the threshold to 5\,NPE further improves sensitivity by approximately 14 times for a 2\,GeV/c$^2$ WIMP mass. 

\begin{figure}[tb!]
    \centering
    \includegraphics[width=1.0\columnwidth]{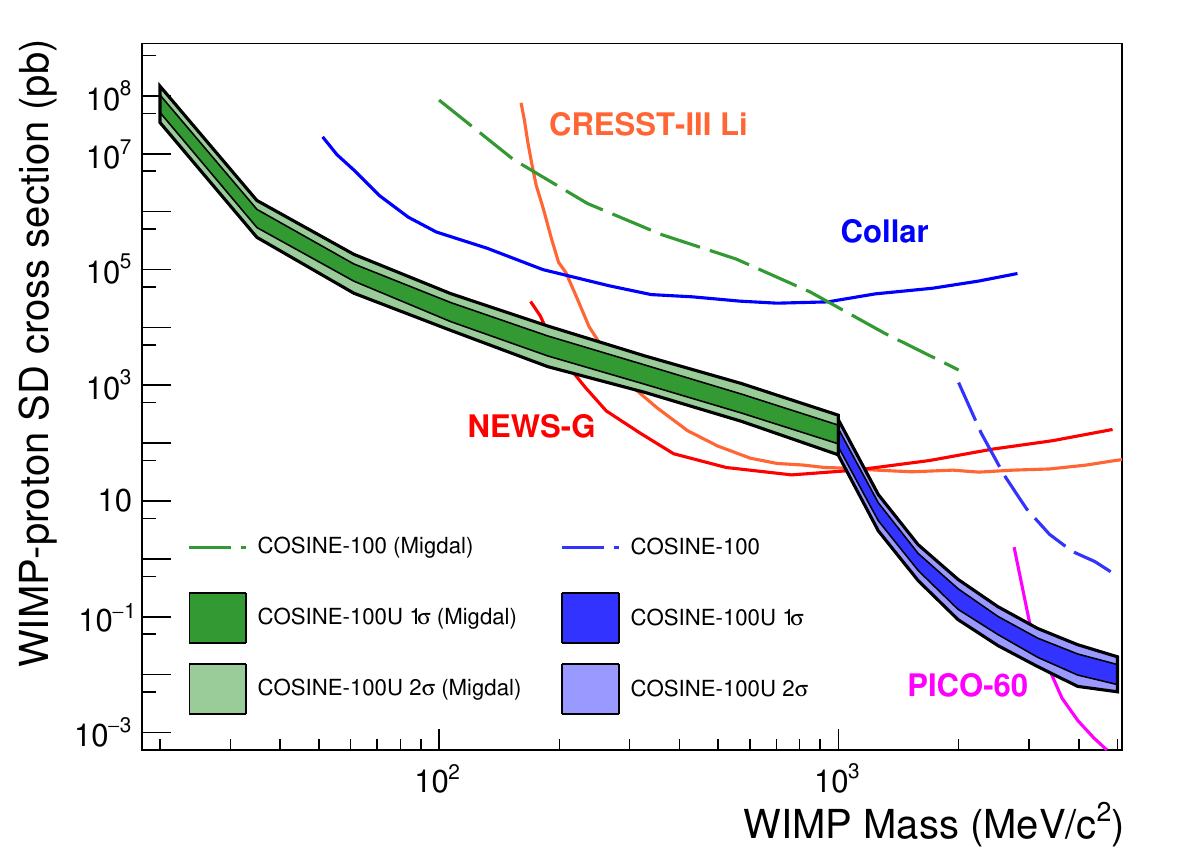}
    
    \caption{
    {\bf Expected sensitivity of the COSINE-100U experiment.}
    The COSINE-100U expected 90\% confidence level limits on the WIMP-proton spin-dependent cross-section, with and without Migdal effect (green and blue bands), are presented assuming the background-only hypothesis indicating the $\pm$1$\sigma$ and $\pm$2$\sigma$ standard deviation probability regions over which the limits have fluctuated. For a comparison, the limits from the COSINE-100 3\,year data~\cite{COSINE-100:2025xqn} are presented.  These limits are compared with the current best limits from CRESST-III Li~\cite{CRESST:2022dtl}, Collar~\cite{Collar:2018ydf}, PICO-60~\cite{PICO:2019vsc}, and NEWS-G~\cite{NEWS-G:2024jms} experiments. 
  }
    \label{fig:COSINE-100U_Sensitivity}
\end{figure}

The COSINE-100U expected limits, assuming a 5\,NPE threshold and the measured light yields of Table~\ref{table:LY} are compared with the current best limits on low-mass WIMP-proton spin-dependent interactions from PICO-60~\cite{PICO:2019vsc}, CRESST-III Li~\cite{CRESST:2022dtl}, NEWS-G~\cite{NEWS-G:2024jms}, and Collar~\cite{Collar:2018ydf} as shown in Fig.~\ref{fig:COSINE-100U_Sensitivity}. They are also compared to COSINE-100 limits from three years of data~\cite{COSINE-100:2025xqn}. Due to sodium's odd-proton numbers and relatively low atomic mass, the COSINE-100U experiment has the potential to explore uncharted  parameter spaces for WIMP masses below 3\,GeV/c$^2$, potentially reaching masses as low as 20\,MeV/c$^2$ when considering the Migdal effect.

\section{Summary}\label{sec7}
The COSINE-100U experiment represents a major upgrade from the COSINE-100 experiment, aimed at improving sensitivity to low-mass dark matter detection. After 6.4\,years of successful operation at the Yangyang Underground Laboratory, the experiment was relocated to the newly constructed Yemilab in Korea, which provides a deeper underground environment with enhanced shielding from cosmic muons. 
Key improvements in the COSINE-100U experiment include the implementation of an improved crystal encapsulation technique that increases light collection efficiency by approximately 35\%. We have evaluated the expected sensitivities of the COSINE-100U experiment, assuming a total mass of 99.1\,kg, a 1-year operation period, and a 5\,NPE analysis energy threshold. Under these conditions, the COSINE-100U detector has the potential to explore previously uncharted parameter spaces for spin-dependent WIMP-proton interactions.

%\backmatter

%\bmhead{Supplementary information}

\acknowledgements
We thank the Yemilab operation team for their dedicated support at Yemilab and the IBS
Research Solution Center (RSC) for providing high performance computing resources. This work was supported by the Institute for Basic Science (IBS) under project code IBS-R016-A1, NRF-2021R1A2C3010989, NRF-
2021R1A2C1013761 and RS-2024-00356960, Republic of
Korea; NSF Grants No. PHY-1913742, DGE-1122492, WIPAC, the Wisconsin Alumni
Research Foundation, United States.

\bibliographystyle{PRTitle}
\providecommand{\href}[2]{#2}\begingroup\raggedright\endgroup

\clearpage

\renewcommand{\thefigure}{A\arabic{figure}}
\renewcommand{\thetable}{A\arabic{table}}
\renewcommand{\theequation}{A\arabic{equation}}
\setcounter{figure}{0}
\setcounter{table}{0}
\setcounter{equation}{0}
\setcounter{section}{0}

\section{Appendix}\label{sec8}
\subsection{Design of COSINE-100U Encapsulation}
The COSINE-100 experiment consists of eight NaI(Tl) crystals, hermetically encapsulated in oxygen-free copper (OFC) enclosures, and submerged in a liquid scintillator veto system to reduce external backgrounds~\cite{Adhikari:2017esn}.

In the NEON experiment, the light collection efficiency was optimized by using 3-inch diameter crystals, which were matched in size to the PMTs, ensuring efficient optical coupling~\cite{NEON:2022hbk}. This approach simplified the encapsulation process and reduced potential photon loss. 

\begin{figure*}[!htb]
  \begin{center}
    \includegraphics[width=1.0\textwidth]{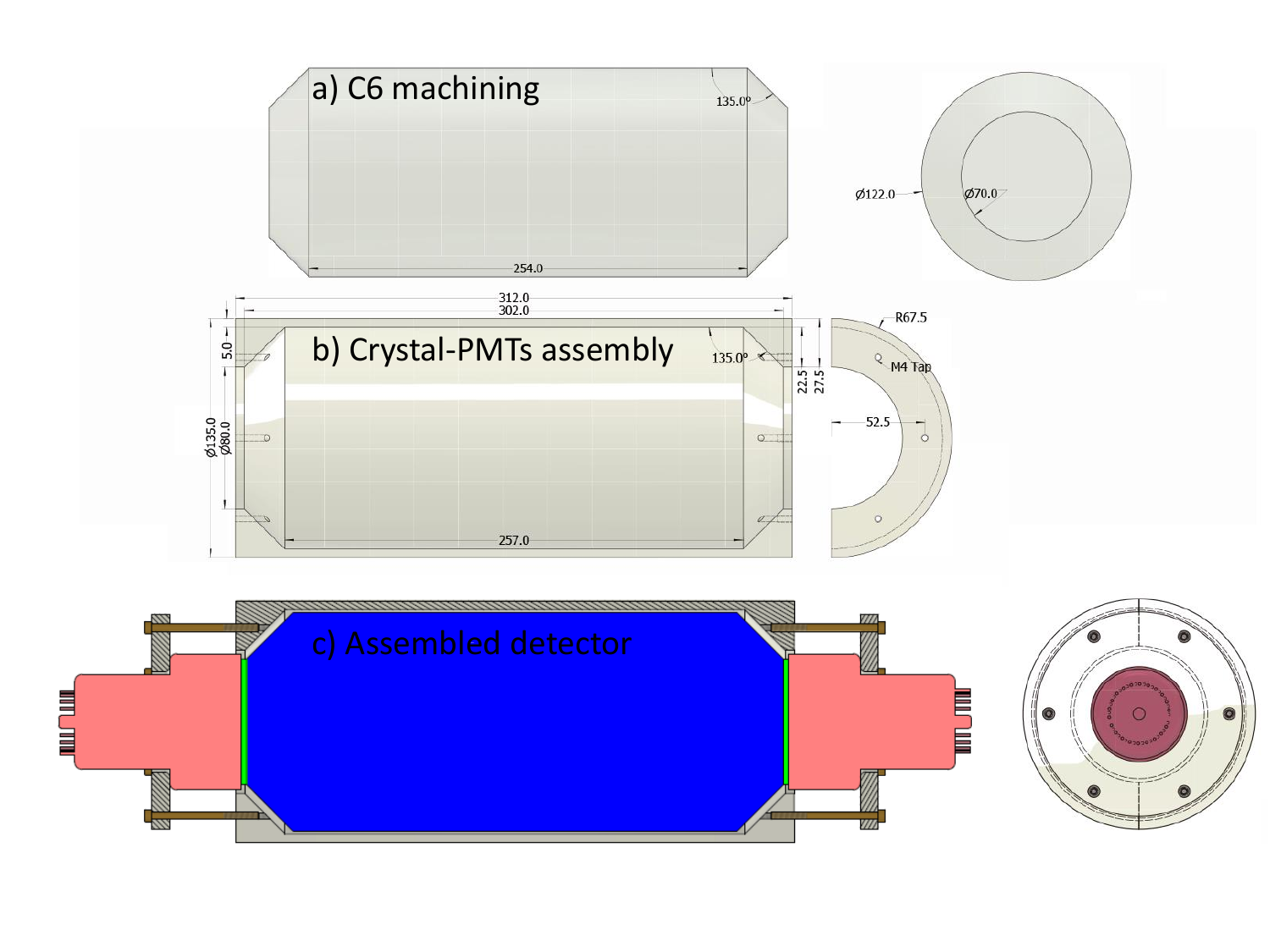}
  \end{center}
  \caption{
{\bf An example of the inner structure for the C6 crystal.} The inner structure was designed to maintain a stable coupling between the PMTs and crystal. 
(a) The design of the C6 machining for guiding light to 3-inch PMTs. 
(b) A PTFE structure that covers the crystal and connects it to the PMTs. 
(c) An illustration of the crystal-PMT assembly, including 2\,mm thick optical pad for coupling. 
    }
\label{fig:cosine100-C6design}
\end{figure*}

However, the COSINE-100 crystals have larger diameters than the PMTs, as summarized in Table~\ref{table:Crystals}, requiring an additional modification to maintain high light collection efficiency.  To address this, the edges of the COSINE-100 crystals were beveled at a 45$^{\circ}$ angle, as shown in Fig.~\ref{fig:cosine100-C6design}(a) for the C6 crystal. This design effectively guides light from the larger crystal diameter to the 3-inch PMTs. 

During this process, the effective area of the 3-inch PMT photocathode, typically 72\,mm, was considered, and the crystal edges were machined to a 70\,mm diameter to optimize PMT attachment. As shown in Table~\ref{table:Crystals}, this machining process results in a mass loss of approximately 1.1\,kg, 0.8\,kg, and 0.5\,kg for 5-inch, 4.8-inch, and 4.2-inch crystals, respectively, primarily due to the 45$^{\circ}$ edge beveling.

To couple the crystal with the PMTs, we designed a 5\,mm thick PTFE structure to surround the crystal, as shown in Fig.~\ref{fig:cosine100-C6design}(b). This PTFE structure is horizontally separable at the center, allowing the crystal to be placed inside. Each end of PTFE structure is connected to the PMT using a PTFE ring with brass bolts, applying pressure to the optical pad between the crystal and the PMT for optical light coupling, as illustrated in Fig.~\ref{fig:cosine100-C6design}(c). 

\begin{figure*}[!htb]
  \begin{center}
    \includegraphics[width=1.0\textwidth]{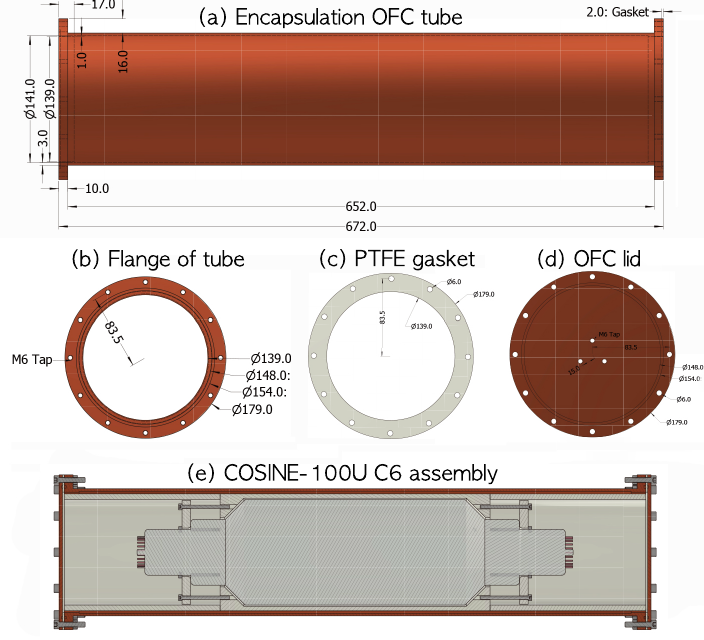}
  \end{center}
  \caption{
{\bf An example of the outer structure for the C6 crystal.} 
The outer structure was designed to prevent the infiltration of outside air and liquid scintillator.  
(a) The OFC tube for the C6. 
(b) Side view of the OFC tube showing the flange with two grooves for tight sealing using gaskets. 
(c) Design of the PTFE gasket. (d) OFC lid, which couples with the flange of the tube through the PTFE gasket.
(e) Illustration of the combined inner and outer structure for C6
    }
\label{fig:cosine100-C6outer}
\end{figure*}

The outer tube, made from OFC, is based on the updated design of the NEON experiment~\cite{Choi:2024trx}, but with an increased diameter and length. As shown in Fig.~\ref{fig:cosine100-C6outer}(a), the C6 tube has a length of 660\,mm, a diameter of 140\,mm, and a thickness of 2\,mm. The side flanges (Fig.~\ref{fig:cosine100-C6outer}(b)) are 20\,mm thick to provide sufficient pressure for a tight seal. Two grooves were machined into each side to accommodate PTFE gaskets or Viton O-ring, as shown in Fig.~\ref{fig:cosine100-C6outer}(c). 

Initially, two OFC lids (Fig.~\ref{fig:cosine100-C6outer}(d)), each 20\,mm thick,  were attached to the tube flanges using twelve brass bolts through PTFE gaskets. However, after low-temperature tests with two crystals, the PTFE gaskets were replaced with Viton O-rings to ensure reliable sealing under such conditions. Each lid includes three holes to accommodate one high-voltage cable and two signal cables for the anode and dynode readouts~\cite{COSINE-100:2018rxe}. Waterproof cable glands secure the lids, preventing the ingress of liquid scintillator and external air through the cable exit holes. The final assembly of the inner and outer structures is illustrated in Fig.~\ref{fig:cosine100-C6outer}(e). 

\subsection{Crystal Machining and Encapsulation}
\begin{figure*}[tb!]
    \centering
    \includegraphics[width=1.0\textwidth]{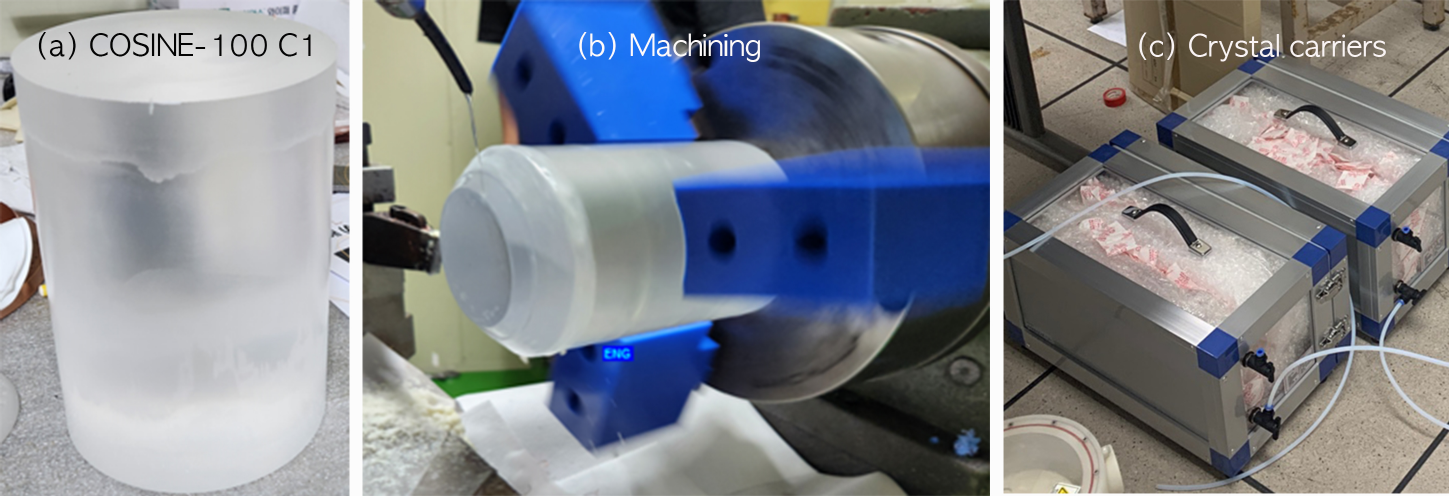}
    \caption{
    {\bf Machining and delivery process.} 
    (a) The COSINE-100 C1 bare crystal after the removal of the original encapsulation. (b) Machining of C1 using a lathe machine. (c) The crystal storage box to transport the machined crystal to glovebox. The storage box was filled with dehumidifiers and flushed with N$_2$ gas to prevent moisture exposure.  }
    \label{fig:C1machining}
\end{figure*}

To upgrade the crystals for the COSINE-100U experiment, the existing encapsulation was first removed, and the crystal edges were machined, as shown in Fig.~\ref{fig:C1machining}. This process was performed by a specialized machining company. A dry room with a dedicated lathe machine was prepared for removing the original encapsulation and shaping the crystal edges to improve light guidance. During machining, mineral oil was continuously poured over the crystal to prevent cracks and suppress NaI(Tl) dust in the environment. Given the highly hygroscopic nature of NaI(Tl) crystals, the machined crystals were stored in a dry storage box and quickly transferred to a low-humidity, N$_2$ gas-flushed glovebox to avoid exposure to atmospheric moisture. 

The humidity level inside the glovebox was maintained at below a few tens of ppm (H$_2$O) using a molecular sieve trap and N$_2$ gas flushing. 
Before polishing the crystals, N$_2$ gas was flushed at a high flow rate of 15 liters per minute for 2 hours, which was then reduced to 5 liters per minute to minimize radon levels inside the glovebox. 

\begin{figure*}[tb!]
    \centering
    \includegraphics[width=1.0\textwidth]{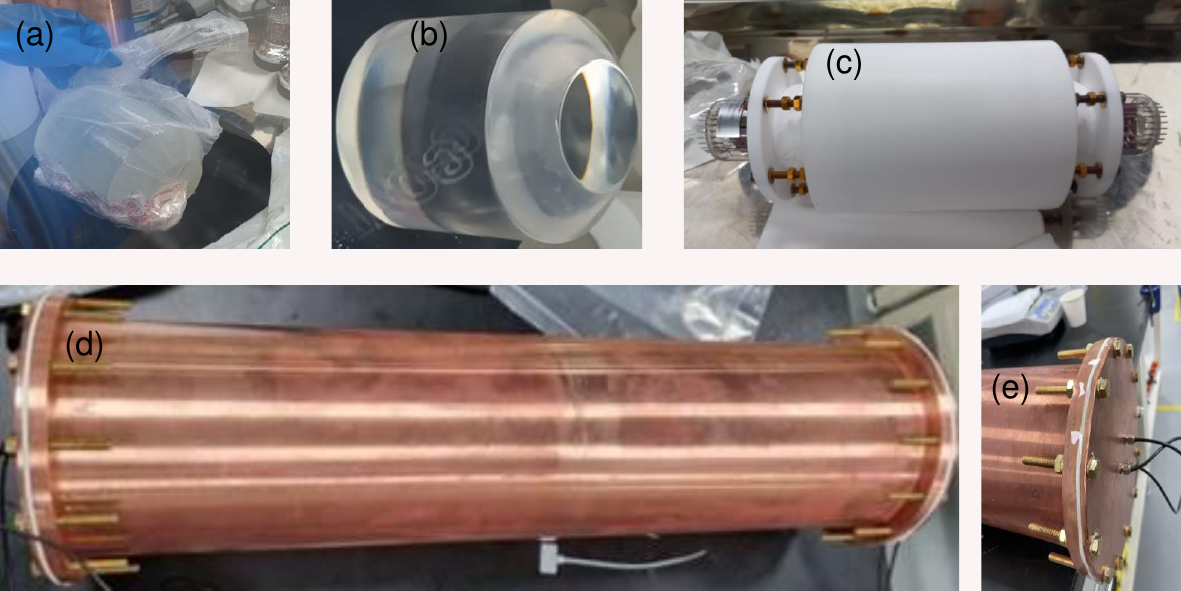}
    \caption{
    {\bf Example of the COSINE-100U crystal encapsulation process.} (a) The C1 crystal was delivered to a low-humidity glovebox after machining for light guidance. (b) The crystal surface was cleaned and polished. (c) The inner structure was assembled by directly attaching the PMTs to the crystal using an optical pad. (d) The inner structure was inserted into the outer OFC tubes and sealed with two OFC lids using PTFE gaskets. (e) The outer lids feature three outlets for signal and high-voltage cables, secured with waterproof cable glands.  }
    \label{fig:encapsulation}
\end{figure*}

After machining, the crystals were contaminated with mineral oil, NaI(Tl) powder, and other debris, as shown in Fig.~\ref{fig:encapsulation}(a). To clean the crystal surfaces, they were gently wiped with anhydrous ethanol.  Due to the small amount of water in the ethanol, this process slightly removed the outer crystal layer. 
Each end of the crystal was then mirror-polished using a polishing pad and SiO$_2$ abrasives in two-steps: first with 3\,$\mu$m and then with 0.5\,$\mu$m particle sizes. After polishing, all surfaces were wiped with cleanroom wipes soaked in anhydrous isopropanol, resulting in a shiny finish on the crystal surfaces,  as shown in Fig.~\ref{fig:encapsulation}(b).  The stability of the polished crystal surfaces was tested in the glovebox for a week, during which no visible changes were observed. 

In parallel with crystal polishing, all encapsulation materials used in the inner and outer structures were cleaned.  Except for the PMTs and crystals, all components were sonicated in a solution of ultrapure water with 5$\%$ Citranox and 5$\%$ oxalic acid for 30\,minutes, twice. After cleaning, the components were dried in a vacuum oven at 120$^{\circ}$C for 12 hours to remove moisture and were then placed in the low-humidity glovebox for more than one day before encapsulation.

The polished crystals were wrapped with 250\,$\mu$m thick Teflon sheets to enhance light collection. 
They were directly attached to the PMTs using 2\,mm thick optical pads, held in place by a  PTFE inner structure with six screw holes, as shown in Fig.~\ref{fig:encapsulation}(c). PTFE pressure rings were used to apply sufficient pressure between the crystal and PMTs for proper optical coupling, secured with brass bolts, nuts, and washers. Special locking washers were used to prevent loosening due to vibrations during transport and operation. The entire assembly was then placed inside a copper case and sealed with copper lids using brass bolts and nuts, as shown in Fig.~\ref{fig:encapsulation}(d). To prevent external contamination, PTFE gaskets and double layers of PTFE reflective sheets were inserted at the flange joints. Cables were routed through cable glands at the center of the copper lid, as shown in Fig.~\ref{fig:encapsulation}(e).

\end{document}